\documentclass[aps,prd,floats,superscriptaddress,showpacs,nofootinbib,twocolumn,preprintnumbers]{revtex4-1}

\usepackage{amssymb}
\usepackage{amsmath}
\usepackage{gensymb}

\usepackage{graphicx}
\usepackage{graphics}
\usepackage{dcolumn}
\usepackage{color}
\usepackage{rotate}
\usepackage{fancyhdr} 
\usepackage{hyperref}
\usepackage{indentfirst}
\usepackage{braket}

\usepackage{umoline}
\usepackage{ulem}

\def\be{\begin{eqnarray}}
\def\ee{\end{eqnarray}}
\def\ba{\begin{eqnarray}}
\def\ea{\end{eqnarray}}
\def\no{\nonumber}

\def\bea{\begin{eqnarray}}
\def\eea{\end{eqnarray}}

\def\bfp{{\bf p}}
\def\bfq{{\bf q}}

\newcommand{\bfk}{\mathbf{k}}

\newcommand{\M}{m_\star}

\definecolor{darkred}{rgb}{.743,0,0}

\newcommand{\refeq}[1]{Eq.~(\ref{eq:#1})}          
          
\newcommand{\reffig}[1]{Fig.~\ref{fig:#1}}          
\newcommand{\refsec}[1]{Sec.~\ref{sec:#1}}
\newcommand{\refapp}[1]{App.~\ref{app:#1}}

\begin{document}
\title{Assessing the Fornax globular cluster timing problem in different models of dark matter}

\author{Nitsan Bar}\email{nitsan.bar@weizmann.ac.il}\affiliation{Department of Particle Physics and Astrophysics, Weizmann Institute of Science,
	Rehovot 7610001, Israel} 

\author{Diego Blas}\email{diego.blas@cern.ch}
\affiliation{Theoretical Particle Physics and Cosmology Group,
	Department of Physics, \\ King's College London, Strand, London WC2R 2LS, United Kingdom}

\author{Kfir Blum}\email{kfir.blum@weizmann.ac.il}\affiliation{Department of Particle Physics and Astrophysics, Weizmann Institute of Science,
	Rehovot 7610001, Israel}

\author{Hyungjin Kim}\email{hyungjin.kim@desy.de}\affiliation{DESY, Notkestrasse 85, 22607 Hamburg, Germany}\affiliation{Department of Particle Physics and Astrophysics, Weizmann Institute of Science,
	Rehovot 7610001, Israel} 
\preprint{KCL-2021-07}
\preprint{DESY 21-023}

\date{\today}

\begin{abstract}
We investigate what the orbits of globular clusters (GCs) in the Fornax dwarf spheroidal (dSph) galaxy can teach us about dark matter (DM). 
This problem was recently studied for ultralight dark matter (ULDM). We consider two additional models: (i) fermionic degenerate dark matter (DDM), where Pauli blocking should be taken into account in the dynamical friction computation; and (ii) self-interacting dark matter (SIDM).  
We give a simple and direct Fokker-Planck derivation of dynamical friction, new in the case of DDM and reproducing previous results in the literature for ULDM and cold DM. 
ULDM, DDM and SIDM were considered in the past as leading to cores in dSphs, a feature that acts to suppress dynamical friction and prolong GC orbits. 
For DDM we derive a version of the cosmological free-streaming limit that is independent of the DM production mechanism, finding that DDM cannot produce an appreciable core in Fornax without violating Ly-$\alpha$ limits. If the Ly-$\alpha$ limit is discounted for some reason, then stellar kinematics data does allow a DDM core which could prolong GC orbits.
For SIDM we find that significant prolongation of GC orbits could be obtained for values of the self-interaction cross section considered in previous works. 
In addition to reassessing the inspiral time using updated observational data, we give a new perspective on the so-called GC timing problem, demonstrating that for a cuspy cold DM profile dynamical friction predicts a $z=0$ radial distribution for the innermost GCs that is independent of initial conditions. The observed orbits of Fornax GCs are consistent with this expectation with a mild apparent fine-tuning at the level of $\sim25\%$. 
%
%
\end{abstract}

\maketitle

\tableofcontents


\section{Introduction}\label{sec:intro}
The Milky Way dwarf spheroidal (dSph) satellite galaxies are broadly believed to be dominated by dark matter (DM) \cite{walker2009universal,Cole2012}\footnote{See, however, a contrary claim in \cite{Hammer:2020qcd}.}, and this fact combined with their small sizes and nearby locations makes them interesting test beds of the small-scale behavior of DM \cite{Flores:1994gz,Moore1994,de2010core,Fattahi:2016nld,Read:2017lvq,Chang:2020rem,Read:2018fxs}. In fact, some of the basic predictions of the most commonly considered paradigm of DM --- collisionless cold dark matter (CDM) --- may be in tension with observations (see, e.g., Refs. \cite{Bullock:2017xww,Salucci:2018hqu}). Conclusive kinematic data for a decisive test of CDM in dSphs is difficult to obtain, but upcoming observatories may supply it \cite{simonLOI}.

\begin{figure}[htbp!]
	\centering
	\includegraphics[width=0.45\textwidth]{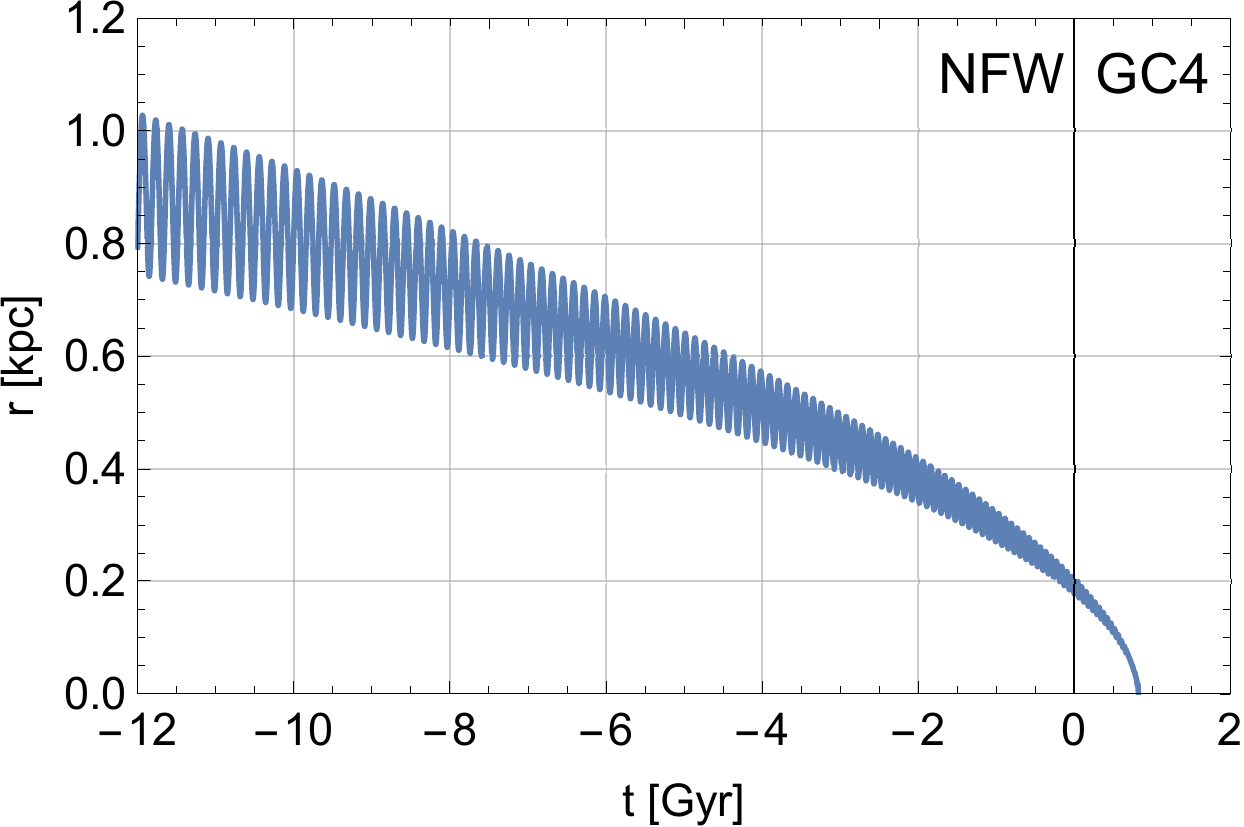}	
	\caption{Orbital radius vs. time, calculated for the Fornax GC4 assuming a slightly eccentric orbit. $ t=0 $ represents today. The orbit calculation assumes the CDM Navarro-Frenk-White (NFW) \cite{Navarro:1996gj} profile of Ref.~\cite{Meadows20}.}\label{fig:gc4long}
\end{figure}

One intriguing puzzle about the dSph galaxies concerns the globular clusters (GCs) of the Fornax dSph \cite{Tremaine1976a}: some of Fornax's six known GCs \cite{Cole2012,wang2019rediscovery} have orbital decay times due to dynamical friction (DF) which seem to fall significantly short of their age \cite{Tremaine1976a}. If estimated na\"ively based on the Chandrasekhar formula \cite{Chandra43}, assuming the usual CDM cusp density profile (see, e.g. Ref.~\cite{Meadows20}), one obtains an instantaneous DF time of less than 1~Gyr for the most troublesome GC4. On the other hand, the stellar content of the GCs is old, $ >10 ~$Gyr \cite{de2016four,Mackey2003a}, as is much of the stellar content of Fornax itself \cite{del2013spatial,wang2019morphology}. 
It may seem unlikely then, that we observe some of the GCs just a short time before they fall to the center of the galaxy. We show a visualization in \reffig{gc4long}.

Part of the scope of this work is to give an analytical perspective on DF, allowing us to sharpen the GC timing puzzle. 
When the dust settles (in Sec.~\ref{ss:stat}, using tools developed throughout the paper) we obtain reasonably robust predictions for the late-time distribution of GCs: for a cuspy halo, the cumulative number of GCs contained within radius $r$ has the form
\be\label{eq:FDtintro} F_{\Delta t}(r)&\propto&\frac{\tau(r)}{\Delta t},\ee
where $\tau(r)$ is the instantaneous DF time (defined precisely in Sec.~\ref{s:cdmcorecusp}), $\Delta t$ is the age of the system, and the prefactor is proportional to the initial number of GCs contained inside $r\sim1$~kpc. Up to the prefactor, the radial slope in Eq.~(\ref{eq:FDtintro}) is insensitive to the assumed initial distribution of GCs and can be calibrated observationally from kinematics modeling. Eq.~(\ref{eq:FDtintro}) substantiates the expectation that the fine-tuning associated with observing a GC at short DF time $\tau\ll\Delta t$ is of order $\tau/\Delta t$. 

We should say in advance that although the timing puzzle is very interesting, our analysis suggests that the possible tension it entails is not very severe. For a cuspy CDM halo, when one takes into account projection effects and the fact that Fornax hosts not just one, but a collection of GCs, then the timing puzzle may be ascribed to a mild (but quite persistent) chance fluctuation with a probability of 25\% or so. 
The lack of a $\sim10^6$~M$_\odot$ nuclear star cluster in Fornax, the remnant of old tidally disrupted GCs~\cite{CapuzzoDolcetta:2008me,CapuzzoDolcetta:2008jy}, may exacerbate the tension. 

Many explanations were suggested for the GC timing puzzle \cite{Hernandez:1998hf,ohlinricher2000,Lotz:2001gz,Goerdt2006,SanchezSalcedo:2006fa,angus2009resolving,Cowsik:2009uk,Kaur2018,Hui2017,Leung_2019,boldrini2020embedding,Berezhiani2019,Hartman:2020fbg,Bar-Or:2018pxz,Lancaster2020}, of which an exciting class of ideas entails a modification to the nature of dark matter, going beyond CDM \cite{Hui2017,Berezhiani2019,Hartman:2020fbg,Bar-Or:2018pxz,Lancaster2020}. In particular, Refs. \cite{Hui2017,Bar-Or:2018pxz,Lancaster2020} studied ultralight dark matter (ULDM) and showed that in the particle mass window $ m\lesssim 10^{-21} $~eV, ULDM would suppress DF enough to eliminate the timing puzzle. However, most of this mass range for ULDM has been scrutinized in the last few years, resulting in disfavoring evidence \cite{Irsic:2017yje,Bar:2018acw,Marsh:2018zyw,Safarzadeh:2019sre}. 
Motivated by the fact that the combination of GC age and orbit measurements probes the details of the DM halo and microphysics, we extend the DF analysis to additional DM models. 
The first model is degenerate dark matter (DDM), in which the phase-space distribution of DM in dSph cores is affected by Pauli blocking \cite{Domcke2015,Randall2017}. The second model is self-interacting DM (SIDM), in which self-interactions between DM particles produce a cored isothermal distribution. 

In Sec.~\ref{sec:DFfermionic} we focus on the microphysics and calculate DF for CDM, DDM, and ULDM. Our results for DDM are new; for ULDM, we make contact with a different derivation in the literature; while for SIDM the microphysics of the DF calculation is argued to be similar to that in CDM. 

All three DM models can, in principle, naturally produce cored isothermal halos. As we show in Sec.~\ref{s:cdmcorecusp}, a cored isothermal distribution of DM suppresses DF in part due to a phase-space effect (associated with the ``core stalling" \cite{Petts2015} identified in past numerical work), as the velocity of the inspiraling GC can become parametrically lower than the DM velocity dispersion. 

For ULDM, DF and the Fornax GC timing puzzle were studied in recent works \cite{Hui2017,Bar-Or:2018pxz,Lancaster2020} and we do not review them again. As noted above, constraints from galaxy dynamics and from cosmological Ly-$\alpha$ analyses suggest a similar behavior to CDM. 

For DDM (Sec.~\ref{sec:ddm}),
we formulate a robust version of the Ly-$\alpha$ bound that is insensitive to DM model building and cosmological history, finding that it disfavors an appreciable core. If one chooses to discount the Ly-$\alpha$ bound (see, e.g. Ref.~\cite{Hui2017} for a qualitative discussion of concerns regarding systematic uncertainties), then stellar kinematics does allow a considerable DDM core which could lead to significant suppression of DF and prolong the settling time of the innermost GCs.

For SIDM (Sec.~\ref{sec:sidm}), stellar kinematics allows a considerable core. If the SIDM cross section is as large as that considered  in Ref.~\cite{Kaplinghat:2015aga}, then the DF settling time for the innermost GCs can be significantly longer than in the cuspy halo CDM model. 

The possibility that baryonic feedback deforms a CDM cusp into a core is also considered. Since baryonic feedback is expected to deform the halo primarily within the half-light radius \cite{Pontzen2012,Oman2016,Meadows20,Read:2018fxs}, the resulting core is spatially smaller than the typical cores that were previously suggested as an explanation to the GC timing puzzle \cite{Goerdt2006,Meadows20}. In that sense, such a model is a hybrid between other cusp/core classes of density profiles that we consider in this work, in the spirit of Ref.~\cite{Cole2012}. As a benchmark, we adopt the density profile fit in Ref.~\cite{Read:2018fxs}. We find that GC orbital decay times may be somewhat prolonged within the inner few hundred parsecs compared to the pure cusp case. This baryon-induced core model may therefore provide a better fit to the GC distribution compared to the cusp case.

Our approach is mostly analytical. Of course, this has limitations and one may be justified in expecting that more progress would require numerical simulations. 
According to recent simulations in Ref.~\cite{Shao:2020tsl}, reasonable initial conditions for the Fornax GCs (derived from the simulations) can lead to the observed configuration in a standard cuspy CDM halo. The timing puzzle may thus be even less significant than the mild 25\% that we find with analytical tools. 
Nevertheless, we believe that analytical insight is important. Notably, as we demonstrate in our analysis, it allows to identify which features of the late-time state of a GC configuration are the result of particular initial conditions and which are generic outcomes of DF.
%

We summarize in Sec.~\ref{sec:summary}. Many details of the calculations are deferred to the Appendices.

\section{Dynamical friction: microphysics}\label{sec:DFfermionic}

Dynamical friction can be described in terms of the Fokker-Planck theory for the motion of a probe particle (a GC in our case) traveling through a gas of spectator particles (DM particles in our case). In App.~\ref{app:DF} we derive the Fokker-Planck equation as the small-momentum-exchange limit of the Boltzmann equation, governing the motion of a probe object in different background media, accounting for the gravitational interaction between the probe and the medium particles. Our calculation is direct, in the sense that it simply amounts to computing the collision integral while taking care to account for the quantum statistics of  spectator gas particles. Here we bypass the details of the calculation, while utilizing the main results. 

The Fokker-Planck equation is characterized by a set of momentum space diffusion coefficients, calculated in App.~\ref{appss:class},~\ref{appss:dfddm}, and~\ref{appss:dfuldm} for the case of a medium composed of a classical gas, degenerate Fermi gas, and Bose gas, respectively. 
Of particular importance for our analysis is the diffusion coefficient $D_{||}$, corresponding to the diffusion in momentum parallel to the probe object's instantaneous velocity. The DF deceleration acting on a probe with mass $\M$ moving with instantaneous velocity ${\bf V}$ w.r.t. the medium is computed as \cite{BinneyTremaine2}
\be
\frac{d\mathbf{V}}{dt} &=& \frac{D_{||}}{\M}\hat{\mathbf{V}}\no\\
&=&-\frac{4\pi G^2\M\rho}{V^3}\,C\,{\bf V}.
\ee
In the second line, to compare the DF arising in different types of media we define the dimensionless coefficient $C$ as follows \cite{Hui2017}:
%
\be C&=& -\frac{V^2 D_{||}}{4\pi G^2\M^2\rho},\ee 
where $\rho$ is the mass density of the medium. 

Different microphysics properties of the medium (in our case, the DM galactic halo) predict different results for $C$. In the next subsection we discuss three scenarios.
%

\subsection{Classical gas}
This is the appropriate limit for a halo composed of a gas of classical particles. We will adopt this limit to describe DF in the ordinary CDM model, as well as for the SIDM model\footnote{This is a good approximation for the SIDM cross-sections of interest, which are small enough such that SIDM particles travel across distances larger than the size of the system without colliding with each other. See Sec.~\ref{sec:sidm}.}. For a homogeneous classical gas with an isotropic distribution function $f_v(v)$, DF is described by the Chandrasekhar formula~\cite{Chandra43} (see also App.~\ref{appss:class}),
\be\label{eq:Chand} C_{\rm class}&=&4\pi\ln\Lambda\int_0^Vdv_mv_m^2f_v(v_m),\ee
where $\ln\Lambda$ is the Coulomb logarithm.
If the gas distribution function is a Maxwellian with velocity dispersion $\sigma$, $f_v(v)=(2\pi \sigma^2)^{-3/2}{\rm exp}(-v^2/(2\sigma^2))$, we have
\be\label{eq:CMax} C_{\rm Max}&=&\ln\Lambda \left(\text{erf}(X)-\frac{2X}{\sqrt{\pi}}e^{-X^2}\right)\\
&\to&\ln\Lambda\begin{cases}
	1 & V\gg \sigma \\ \frac{\sqrt{2}}{3\sqrt{\pi}}\frac{V^3}{\sigma^3} & V\ll \sigma
\end{cases},\no\ee
where $X\equiv V/(\sqrt{2}\sigma)$ and where in the second line we show the asymptotic scaling of $C$ at large and small $X$.

\subsection{Degenerate Fermi gas}\label{ss:ddm}
This is the relevant limit for DF at the core of a halo supported by the degeneracy pressure of light fermionic DM (DDM model \cite{Domcke2015,Randall2017}). In the high-degeneracy limit we have $f_v(v)=3/(4\pi v_F^3)\theta\left(v_F-v\right)$, where $\theta(x)$ is the Heaviside function, the Fermi velocity $v_F$ is related to the medium density via
\be\label{eq:rhoddm}\rho&=&\frac{g m^4v_F^3}{6\pi^2},\ee
$m$ is the mass of the particles and $g$ is the number of degrees of freedom (e.g. $g=2$ for Weyl fermions).
The calculation in App.~\ref{appss:dfddm} gives the following limiting behavior,
\be\label{eq:Cdeg} C_{\rm DDM}&\to&\ln\Lambda\begin{cases}
	1 & V\gg v_F \\ \frac{V^3}{v_F^3} & V\ll v_F
\end{cases}.\ee

Thus, in both limits $V\gg v_F$ and $v\ll v_F$, we find that DF in a degenerate medium is equivalent to DF in a classical medium with the replacement $\sigma\to\left(\frac{2}{9\pi}\right)^{\frac{2}{3}}v_F\approx0.17v_F$. 
Note that the three-dimensional velocity dispersion associated with the classical isotropic Maxwellian distribution is $\langle v_x^2+v_y^2+v_z^2\rangle=\langle v^2\rangle=3\sigma^2$, while the dispersion for the degenerate distribution is $\langle v^2\rangle=(3/5)v_F^2$. Therefore, the pressure in the different types of media matches when $v_F\approx2.2\sigma$. Similarly, Eqs.~(\ref{eq:Cdeg}) and~(\ref{eq:CMax}) tell us that DF in these media match when $v_F\approx5.8\sigma$. We note that the form of \refeq{Cdeg} agrees with the results of Ref.~\cite{Chavanis2020landau}\footnote{We thank P.H. Chavanis for pointing it out to us.}

As an aside, it is interesting to note that to leading order in $ m/\M $, the diffusion coefficient of a classical gas has the same functional form with respect to the distribution function as the diffusion coefficient of a degenerate gas (c.f. \refeq{diffclas} and \refeq{diffdeg}). 
This is somewhat surprising, because the Fokker-Planck calculation took into account Pauli exclusion in the medium whereas Eq.~(\ref{eq:Chand}) does not. Moreover, according to Eq.~(\ref{eq:Chand}), only particles with velocities smaller than the probe object's contribute to the DF. For the case of degenerate matter, one could have expected that the opposite should happen: only particles close to the Fermi surface contribute to DF. We refer the reader again to App.~\ref{appss:dfddm} for the detailed computation that leads us to Eq.~\eqref{eq:Cdeg}.

Finally, note that above we evaluated DF in the zero-temperature limit and not in the finite-temperature limit. In \refsec{ddm} we consider a finite-temperature density profile, so we should keep this caveat in mind. We have not explored DF of degenerate matter within the more sophisticated treatment of Refs. \cite{Tremaine1984,Weinberg1986}.

\subsection{Bose gas}
This is the relevant limit for the case where halo particles follow the Bose-Einstein statistics, as in the ULDM model. 
The diffusion coefficients can be obtained either by solving a Langevin equation with stochastic fluctuations of the gravitational potential~\cite{Bar-Or:2018pxz} or, as we do in App.~\ref{appss:dfuldm}, by using a kinetic equation\footnote{While this paper was being prepared for publication, Ref.~\cite{Bar-Or:2020tys} appeared which also presents a kinetic theory derivation of the ULDM diffusion coefficients.}.
Both approaches provide identical results.

Up to a slight modification of the Coulomb logarithm, DF for the bosonic gas includes a contribution to the $C$ term that is identical to that of the classical gas in Eq.~\eqref{eq:CMax}. 
In addition to this, ULDM large-scale density fluctuations (manifested by Bose-enhancement terms in the kinetic theory computation) cause additional velocity drift that can be characterized by an extra term to $C\to C+\Delta C$, with\footnote{Formally, the $\Delta C$ term is there also for standard CDM but is negligible unless the individual DM particles are extremely massive.}
\bea
\Delta C &=& \ln \Lambda \left( \frac{m_{\rm eff}}{m_\star} \right) 
\bigg( {\rm erf}(X_{\rm eff}) - \frac{2 X_{\rm eff}}{\sqrt{\pi}} e^{-X_{\rm eff}^2} \bigg),
\eea
where $m_{\rm eff} = \pi^{3/2} \rho / (m\sigma)^3$ is the ULDM mass enclosed in an effective de Broglie volume and $X_{\rm eff} \equiv v/\sqrt{2} \sigma_{\rm eff}$ with $\sigma_{\rm eff} = \sigma/\sqrt{2}$. Numerically, $ m_{\rm eff}\approx 1.2\times 10^{6}\left(10^{-21}~{\rm eV}/m\right)^3[\rho/(3\times 10^{7}~M_{\odot}/{\rm kpc}^3)][(10~{\rm km/s})/\sigma]^3$~M$_\odot$. With these numbers and keeping in mind a typical GC mass $m_*\sim10^5$~M$_\odot$, the $\Delta C$ effect becomes quantitatively important in Fornax for $ m\lesssim 3\times 10^{-20}$~eV.

The kinetic theory result summarized above assumed that the scale size of the system -- e.g., the radius $r$ of a GC orbit -- is much larger than the effective de Broglie wavelength of the ULDM particles, 
\be r_{\rm dB}&\approx&\frac{2\pi}{m\sigma}\;\approx\;300\left(\frac{10~\rm km/s}{\sigma}\right)\left(\frac{10^{-21}~\rm eV}{m}\right)~{\rm pc},\ee
and thus much larger than ULDM quasi-particle excitations or than the soliton core that is ubiquitously found in ULDM simulations (see Ref.~\cite{Hui2017} for a review). For $r<r_{\rm dB}$, the treatment above breaks down and must be modified by taking into account large-scale coherence effects of the ULDM. This can be done via solving the Schr\"odinger equation, as shown in Refs.~\cite{Hui2017,Lancaster2020}, which indeed found that DF becomes suppressed at $r\lesssim r_{\rm dB}$. We refer the reader to Refs.~\cite{Hui2017,Lancaster2020} for more details on DF and the Fornax GC puzzle in the context of ULDM. Here we only note that for $m\gtrsim10^{-20}$~eV, where $r\gg r_{\rm dB}$ and $m_{\rm eff}\ll m_*$ for  the Fornax GCs, DF in the ULDM medium becomes quantitatively similar to DF in a classical medium.

\section{Dynamical friction in a CDM halo: core vs. cusp}\label{s:cdmcorecusp}
It is natural to define an instantaneous DF time, $\tau$, via
\be\label{eq:genDecaytime2}\tau&=&\frac{V^3}{4\pi G^2\M\rho C},\ee
such that (including here only the DF effect) 
\be{\dot{\bf V}}&=&-\frac{1}{\tau}{\bf V}.
\ee 
A crude estimate of the time scale it would take a GC to settle down to the dynamical center of a halo can be obtained by computing $\tau$, using the current instantaneous position and velocity of the GC. 
Assuming a CDM NFW distribution, and plugging an estimate of the dark matter density and velocity dispersion corresponding to the present observed position of each GC into Eqs.~(\ref{eq:CMax}) and~(\ref{eq:genDecaytime2}), the result we find is summarized in the column marked $\tau_{\rm CDM}$ (highlighted in blue) in Tab.~\ref{tab:dat}. For GC3 and GC4 the DF time estimated in this way is 2.6 and 0.9~Gyr, respectively, much shorter than the age of the system.\footnote{Our estimates are larger than those previously obtained in Ref.~\cite{Hui2017}; we discuss the differences in Sec.~\ref{sec:discussion}.  }
\begin{table*}[htb!]
	\centering
	\caption{Some details of Fornax GCs. For the galactic center of Fornax we use an updated measurement \cite{wang2019morphology}, based on surface brightness modeling. This estimate is $ \approx 160 $~pc off relative to the center defined by previous works \cite{Mackey2003a,Cole2012,Hui2017,Meadows20,boldrini2020embedding,Shao:2020tsl}, 
	leading to different projected radii of GCs. We set the distance to Fornax as $ 147\pm 4 $~kpc \cite{de2016four}. We estimate the error on $ r_\perp $ by propagating the distance error, added in quadrature with a $ 13 $~pc \cite{wang2019morphology} uncertainty on the center. For relative radial velocities $ \Delta v_r $, we use the galactic radial velocity ${\rm RV}_{\rm Fornax}= 55.46\pm 0.63 $~km/s \cite{Hendricks_2014} and set $ \Delta v_r ={\rm RV}_{\rm GC}-{\rm RV}_{\rm Fornax}$, adding errors in quadrature. 
	For GC6, the values correspond to a small sample of stars, likely contaminated by background \cite{wang2019rediscovery}. 
	$ r_{c/h} $ refers to the King radius for GC1-GC5 and half-light radius for GC6. 
	The CDM instantaneous DF time (\refeq{genDecaytime2}) estimates are based on the NFW profile of \cite{Meadows20}. The instantaneous DF times of DDM and SIDM are based on Secs.~\ref{sec:ddm} and~\ref{sec:sidm}.} \label{tab:dat}
	\begin{tabular}{ c| c c c c c| c| c c}
		& $ \M\;[10^5M_{\odot}]$ & $ r_\perp [{\rm kpc}]$  & $ \Delta v_{r} [{\rm km/s}] $ & $ r_{c/h}\;[{\rm pc}] $ & Refs. & {\color{blue}$ \tau_{\rm CDM}~[{\rm Gyr}] $} & $ \tau_{{\rm DDM}}^{(135)}~[{\rm Gyr}] $ & $ \tau_{{\rm SIDM}}~[{\rm Gyr}] $ \\ \hline
		GC1 & $ 0.42\pm 0.10 $ & $ 1.73\pm 0.05 $ & $ 3.54\pm 1.18 $ & $ 10.8\pm 0.3 $ & \cite{de2016four,Lauberts1982,Letarte2006,Mackey2003a,Hendricks_2014} & {\color{blue}$ 119 $} & $ 122 $ & $ 79.3 $\\
		GC2 & $ 1.54\pm 0.28 $ & $ 0.98\pm 0.03 $ & $ 3.9\pm 0.7 $ & $ 6.2\pm 0.2 $ & \cite{de2016four,Morrison2001,Letarte2006,Mackey2003a} & {\color{blue}$ 14.7 $} & $ 7.12 $ & $ 8.82 $\\
		GC3 & $ 4.98\pm 0.84 $ & $ 0.64\pm 0.02$ & $ 4.94\pm 0.66 $ & $ 1.7\pm 0.1 $ & \cite{de2016four,Skrutskie2006,Larsen_2012,Mackey2003a} & {\color{blue}$ 2.63 $} & $ 1.48 $ & $ 2.21 $\\
		GC4 & $ 0.76 \pm 0.15 $ & $ 0.154\pm 0.014 $ & $ -8.26\pm 0.64 $ & $ 1.9\pm 0.2 $ & \cite{de2016four,Skrutskie2006,Larsen_2012,Mackey2003a} & {\color{blue}$ 0.91 $} & $ 10.7 $ & $ 14.8 $\\
		GC5 & $ 1.86\pm 0.24 $ & $ 1.68\pm 0.05 $ & $ 3.93\pm 0.77 $ & $ 1.5\pm 0.1 $ & \cite{de2016four,Skrutskie2006,Larsen_2012,Hendricks_2014,Mackey2003a} & {\color{blue}$ 32.2 $} & $ 30.1 $ & $ 20 $ \\
		GC6 & $ \sim 0.29 $ & $ 0.254\pm 0.015 $ & $ -1.56\pm1.36 $ & $ 12.0\pm 1.4 $ & \cite{wang2019rediscovery,Shao:2020tsl} & {\color{blue}$ 5.45 $} & $ 16.1 $ & $ 22 $ 
	\end{tabular}
\end{table*}

However, estimating an orbital decay time from the instantaneous value of $\tau$ can be misleading. In a realistic galaxy, the DM phase-space distribution and with it the instantaneous value of $\tau$ could change along the orbit of the GC. To obtain a better estimate of the actual settling time one could track the orbit of the GC semi-analytically, using the phase-space-dependent value of $\tau$ along the orbit \cite{Inoue2009,Petts2015,Kaur2018}. Some details of this calculation are given in App.~\ref{app:orbits}. 

\begin{figure}[htbp!]
	\centering
	\includegraphics[width=0.45\textwidth]{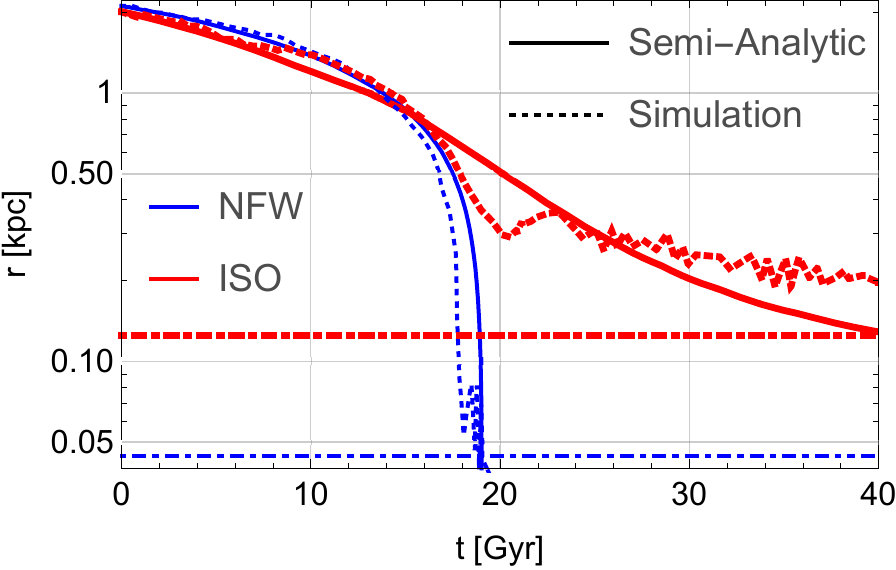}		
	\caption{Radius of an infalling GC with mass $ \M=3\times 10^{5}~M_{\odot} $, based on the simulations of Ref.~\cite{Meadows20}. In dotted blue (thick dotted red) we plot the simulation result (Fig.~3 in Ref.~\cite{Meadows20}) for the NFW (ISO) halo. In solid lines, we plot our semi-analytic integration. 
	Horizontal dot-dashed lines show the radii $ r_{f} $ where $ M_{\rm halo}(r_f)= \M $, in which the semi-analytic treatment should break down.}\label{fig:rvst}
\end{figure}

The semi-analytic integration reproduces results from $ N $-body simulations \cite{Meadows20}.
To demonstrate this, we use Eqs.~\eqref{eq:CMax} and~(\ref{eq:genDecaytime2}) while reading the CDM density and velocity dispersion from the N-body simulations of Ref.~\cite{Meadows20} to integrate the orbit of a GC. In Fig.~\ref{fig:rvst} we compare our results to two different scenarios from \cite{Meadows20}. 

The first scenario, denoted NFW, contains a cuspy NFW-like halo (the density profile of this model is shown in Fig.~\ref{fig:comparemodels}). The orbit of a GC in this halo is shown by the blue dashed line for the simulation of \cite{Meadows20} and by a blue solid line for the semi-analytic tracking.  
The second scenario, denoted ISO, contains an isothermal core halo (also shown in Fig.~\ref{fig:comparemodels}). The GC orbits are shown by the red lines. Again, the semi-analytic method (solid) compares reasonably well with the simulation (dotted). 

The results have a mild dependence on the choice of the Coulomb log, and we make slightly different choices for the different scenarios. For NFW we follow Ref.~\cite{Goerdt2006} in setting
\be
\ln\Lambda_{\rm NFW} &=& \ln \frac{b_{\rm max}\sigma^2}{G\M}.
\ee
However, instead of the $ b_{\rm max}=0.25 $~kpc used in Ref.~\cite{Goerdt2006}, we adopt $ b_{\rm max}=0.5 $~kpc. For ISO we follow \cite{Petts2015,Hui2017}
\be\label{eq:clogISO}
\ln \Lambda_{\rm ISO} = \ln \frac{2V^2r}{G\M} \; .
\ee
We have checked that changing the definition of the Coulomb log according to different prescriptions in the literature changes the predicted infall time of GCs at the level of a few tens of percent. This would not be crucial for our main results.

\begin{figure}[htbp!]
	\centering
	\includegraphics[width=0.45\textwidth]{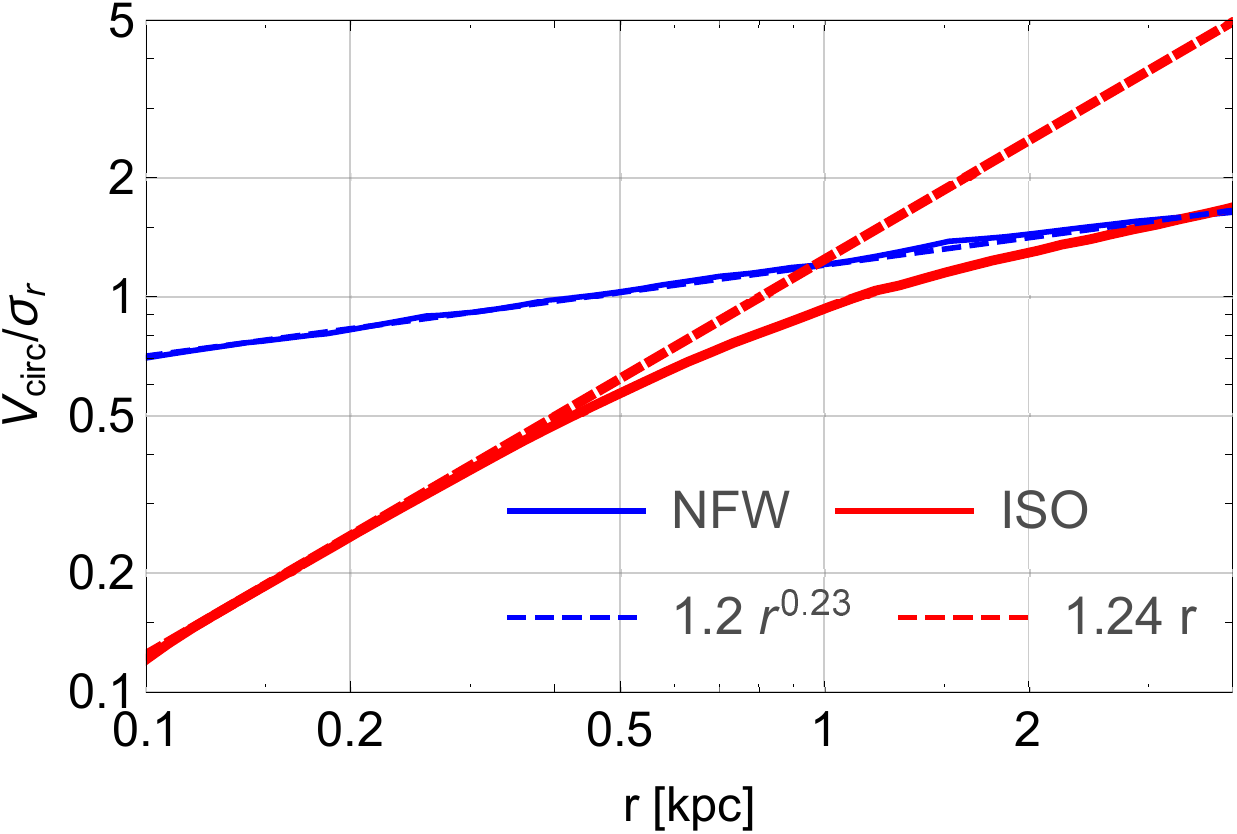}
	\caption{The ratio of the circular velocity $ V_{\rm circ} $ to the radial velocity dispersion $ \sigma_r $, reproduced from Ref.~\cite{Meadows20} for NFW (thick blue) and isothermal (ISO, red) halos. 
	}\label{fig:velratioMeadows}
\end{figure}

\begin{figure}[htbp!]
	\centering
	\includegraphics[width=0.45\textwidth]{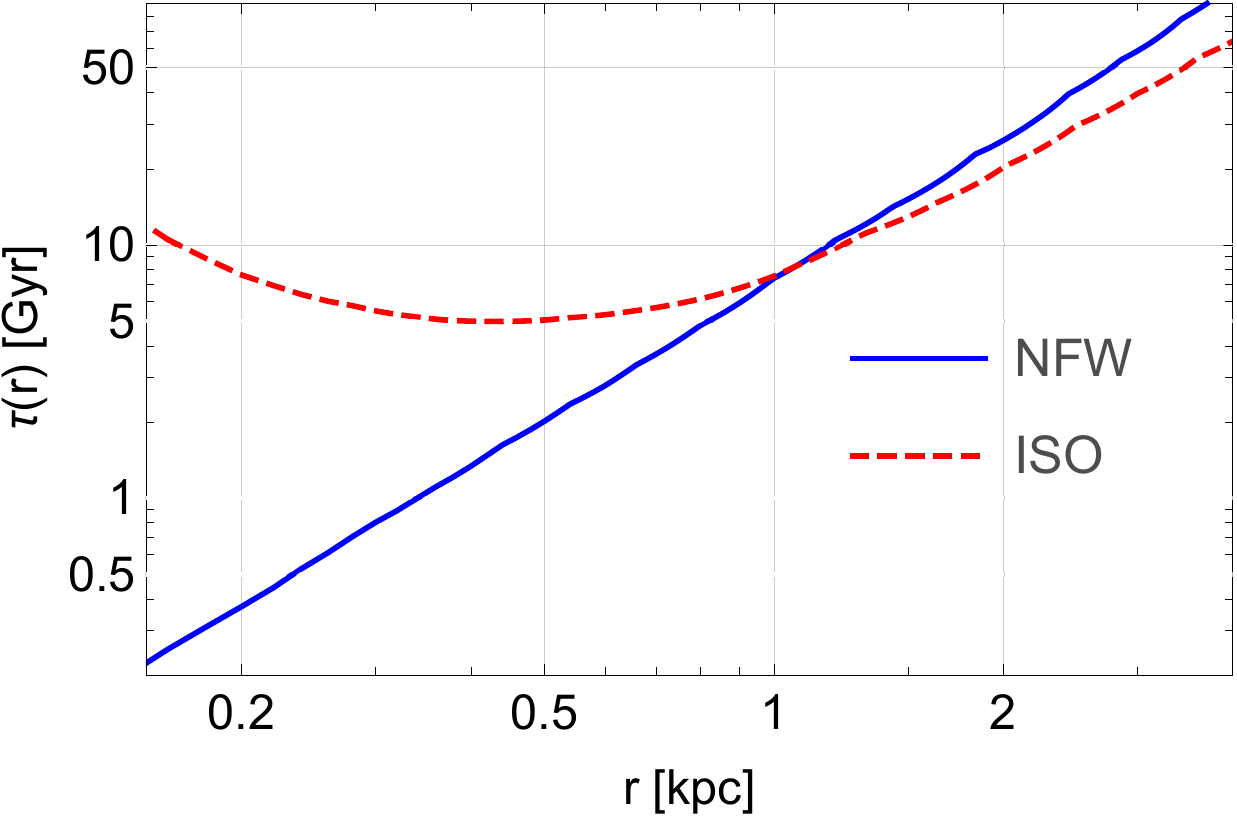}		
	\caption{Orbital decay time calculated using $\rho$ and $\sigma$ from Fig.~1 of Ref.~\cite{Meadows20} and assuming a test object on a circular orbit. NFW denotes a cuspy profile, and ISO denotes an isothermal cored profile. We use here the GC mass $ \M = 3\times 10^5 ~M_{\odot}$.}\label{fig:cdmcore}
\end{figure}

We can gain some insight on the difference between the DF settling time in the cusp vs. the core profiles. 
In the central part of a cuspy NFW halo, the density scales as $ \rho\propto 1/r $ and the circular velocity scales as $ V_{\rm circ} \propto r^{1/2} $. Let us simplify matters by assuming (as was often done in previous works) that the GC moves on an approximately circular orbit, $ V = V_{\rm circ} $. With this, considering the NFW halo of Ref.~\cite{Meadows20} and using \refeq{CMax}, we find\footnote{Ref.~\cite{Hui2017} assumed that $V = V_{\rm circ}$ and also took $V_{\rm circ}$ equal to the velocity dispersion $\sigma$, which would lead to a constant $C_{\rm Max}$. While this is roughly correct, for the NFW halo of Ref.~\cite{Meadows20} we find mild radial dependence of $ V_{\rm circ}/\sigma \propto r^{0.23} $, as shown in \reffig{velratioMeadows}.} $ C_{\rm Max}\approx  0.3(r/{\rm kpc})^{0.5} \ln\Lambda $. The DF time $ \tau $ defined in \refeq{genDecaytime2} then scales as $ \tau\propto r^{2} $. This is a rough estimate: if we use the simulation data of Ref.~\cite{Meadows20} for $\sigma$ and $\rho$ we find a similar but slightly different scaling, $ \tau \propto r^{1.85} $, plotted in solid blue in \reffig{cdmcore}. The important point is the approximately power law decline of $\tau$ towards small $r$. This is the cause of the fast orbital decay of the GC in the cuspy halo model.

The situation is different in a cored halo. In a core, the density $ \rho\approx\rho_0 =  $~const., the circular velocity $V_{\rm circ}\propto r$, while a Jeans analysis shows that for an isotropic velocity distribution the  velocity dispersion is constant $ \sigma \approx \sqrt{G\rho_0}r_c $ \cite{Petts2015}, where $r_c$ is the core radius (see App.~\ref{app:cdm}). 
This implies $ V_{\rm circ}/(\sqrt{2}\sigma) \sim r/r_{c} $, as corroborated in Fig.~\ref{fig:velratioMeadows} by comparing to the simulation data from \cite{Meadows20}. 
At $r<r_c$ the low-velocity approximation in Eq.~\eqref{eq:CMax} gives $ C \propto (r/r_c)^3\ln\Lambda $. The $ (r/r_c)^3 $ factor can be thought of as a phase-space suppression of DF: it arises from the factor $ \int_0^{V}dv_m v_m^2f_v(v_m) $ in Eq.~(\ref{eq:Chand}), because the velocity dispersion inside an isotropic core is greater than the circular velocity (which we assumed to match the instantaneous GC velocity). 

Altogether, referring to Eq.~\eqref{eq:genDecaytime2}, an isotropic core predicts an approximately constant $\tau$ (see also Ref.~\cite{Hernandez:1998hf}),
\be\label{eq:MaxwellianTauDFLowV}
\tau & \approx & \frac{3\sqrt{\pi}}{\sqrt{2}}\frac{\sigma^3}{4\pi G^2 \M\rho_0\ln\Lambda } \\ & \approx & \frac{\pi}{2\sqrt{3}} \frac{r_c^3\sqrt{\rho_0}}{\sqrt{G} \M \ln \Lambda} \no \\ 
%
%
& \approx & 1.95\frac{4}{\ln\Lambda}\left(\frac{r_c}{1~{\rm kpc}}\right)^3\frac{3\times 10^5~M_{\odot}}{\M}\left(\frac{\rho_0}{3\times 10^7\frac{M_{\odot}}{{\rm kpc}^3}} \right)^{\frac{1}{2}} {\rm Gyr} \; . \no
\ee
In the second line we used Eq.~\eqref{eq:sigrFC} and in the third line we used values relevant for Fornax GCs. 
Again, we can compare this estimate to numerical simulations. The dashed red line in Fig.~\ref{fig:cdmcore} shows $\tau$ as calculated by using the velocity dispersion and density read off the cored ISO model of Ref.~\cite{Meadows20}. $\tau$ is approximately constant in the core, exceeding the value of $\tau$ found for the NFW halo. 

The result that DF is suppressed in a cored halo, in comparison to a cusp, is consistent with the finding of Refs.~\cite{Goerdt2006,Read:2006fq,Cole2012}, further confirmed in Refs.~\cite{Inoue2009,Petts2015,Kaur2018,Meadows20}.\footnote{\label{fn_N}The core stalling observed in $ N $-body simulations was initially ascribed in Ref.~\cite{Read:2006fq} to a failure of the Chandrasekhar formula. However, as we explained here (see also Ref.~\cite{Petts2015}), semi-analytic tracking using the Chandrasekhar formula along the orbit reproduces this result.} 
In the next two sections we consider this perspective in exploring DM models that predict a cored halo.

\section{Degenerate dark matter (DDM)}\label{sec:ddm}

Ref. \cite{Domcke2015} (see also \cite{Randall2017}) made the interesting observation that light fermionic DM would produce a core in dSphs, if the DM particle mass $m$ is light enough to place the halo in the degenerate regime. We call this model degenerate DM (DDM).
The DDM core scale radius $r_c$ can be estimated via
\be\label{eq:rcddm} r_c&=&\frac{A}{G^{\frac{1}{2}}\rho_0^{\frac{1}{6}}\left(g\,m^4\right)^{\frac{1}{3}}}\\
&\approx&681\left(\frac{\rho_0}{10^{7}~\rm M_{\odot}/kpc^3}\right)^{-\frac{1}{6}}\left(\frac{g\,m^4}{2\times(120~\rm eV)^4}\right)^{-\frac{1}{3}}~{\rm pc},\no\ee
where $\rho_0$ is the core central density and where $A=\left(9\pi/2^7\right)^{1/6}\approx0.78$. In App.~\ref{app:profile} we give a derivation of Eq.~(\ref{eq:rcddm}), modeling the dSph halo by a maximum entropy configuration (at fixed total mass and energy, similarly to Ref.~\cite{lyndenbell68})\footnote{Our approach in App.~\ref{app:profile} is similar to that of Ref.~\cite{Domcke2015}, but differs in that we include also non-zero temperature solutions. Such solutions were noted but not implemented in Ref.~\cite{Domcke2015}. We find that these solutions could expand the range of applicability of the DDM model in dSphs. Ref.~\cite{Chavanis:2014xoa} also considered non-zero temperature solutions, albeit without comparison to data (We thank P.H. Chavanis for pointing this out to us.) }. The maximum entropy halo is isothermal, scaling as $\rho\propto 1/r^2$ at large $r$.
Between the degenerate core and the $ 1/r^2 $ regime there are intermediate features that depend on the temperature.

Inside $r\lesssim r_c$ DDM particles are described by a degenerate distribution function with Fermi velocity $v_F$ related to their mass density via Eq.~(\ref{eq:rhoddm}). DF for this system is characterized by Eq.~(\ref{eq:Cdeg}), so inside the DDM core, where $\rho$ and $v_F$ are constant,  Eq.~\eqref{eq:genDecaytime2} yields a constant DF time,
\be\label{eq:tauDDMlowvel}
\tau_{\rm DDM} & \approx & \frac{3\pi}{2 G^2 g m^4 \M\ln\Lambda} \\ &= &4.8\frac{4}{\ln\Lambda}\frac{10^5~M_{\odot}}{\M}
\frac{2\times (150~{\rm eV})^4}{gm^4}~{\rm Gyr} \; .\no
\ee
Note that if one inserts the DDM halo core radius (\ref{eq:rcddm}) into Eq.~\eqref{eq:MaxwellianTauDFLowV}, one obtains the same parametric dependence as in Eq.~\eqref{eq:tauDDMlowvel}. 
This is a result of the similarity between Eqs.~(\ref{eq:CMax}) and~(\ref{eq:Cdeg}). The  interesting feature of the DDM model is that it produces the core due to Pauli blocking.

Naively, Eq.~(\ref{eq:tauDDMlowvel}) suggests that DF in a DDM core could be arbitrarily suppressed by decreasing $ m $. This happens because decreasing $m$ at fixed $\rho$ is tied to increasing $v_F$. The DF effect on Fornax GCs is thus an interesting test bed of DDM, and in Sec.~\ref{sec:decay} we explore this point in more detail. Before entering that discussion, however, we first consider observational limits on $m$. 

First, the Fermi velocity cannot be arbitrarily high in a gravitationally bound halo \cite{Tremaine:1979we}. In Sec.~\ref{ss:ddmlosvd} we make this analysis more precise by fitting stellar line-of-sight velocity distribution (LOSVD) data to the DDM halo model; we find that while the Fornax LOSVD data indeed constrains $m\gtrsim100$~eV or so, this constraint by itself would still allow a significant modification of DF compared to the CDM prediction. 

A second and much tighter constraint comes from cosmological structure formation as observed through Ly-$\alpha$ forest statistics. We show in Sec.~\ref{ss:ddmcosmo} that this constraint directly affects the same combination, $gm^4$, that appears in Eqs.~(\ref{eq:tauDDMlowvel}) and~(\ref{eq:rcddm}). Imposing the Ly-$\alpha$ constraint excludes DDM from making an appreciable core in Fornax on scales $r\gtrsim100$~pc, meaning that DDM could not significantly affect the orbits of GCs. While earlier work on DDM argued that a non-thermal production mechanism for DDM could avoid the cosmological constraint, we formulate a rather robust version of the bound which appears difficult to evade.

\subsection{Stellar LOSVD constraints on DDM in Fornax}\label{ss:ddmlosvd}
\begin{figure}[htbp!]
	\centering
	\includegraphics[width=0.45\textwidth]{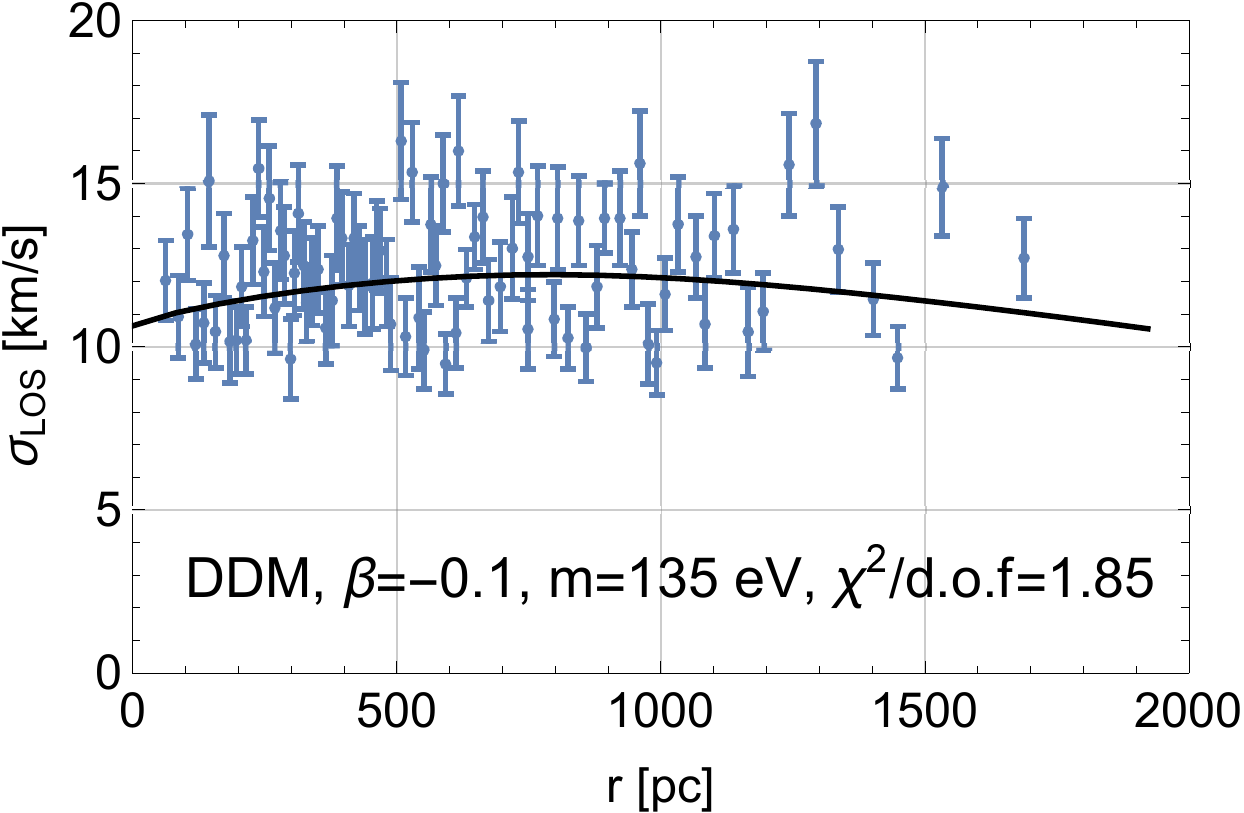}	
	\caption{LOSVD compared with data \cite{Read:2018fxs} for DDM with $g=2$. The best-fit parameters of the profile are $ \mu_0/T=3 $ and $ \rho_0=3.9\times 10^{7}~M_{\odot}/{\rm kpc}^3 $.}\label{fig:losvd}
\end{figure}

In this section we summarize the results of a Jeans analysis for the DDM model in Fornax\footnote{For previous analyses, see Refs.~\cite{Domcke2015,Randall2017,DiPaolo2018,Savchenko2019,Boyarsky2009,Alvey:2020xsk}.}. The DDM profile is described in App.~\ref{app:profile} and the details of the Jeans analysis are given in App.~\ref{app:jeans}. 
In Fig.~\ref{fig:losvd} we plot LOSVD data of Fornax \cite{Read:2018fxs} alongside fits of the density profile presented in App.~\ref{app:profile}. Our fitting procedure is based on a simple $\chi^2$ minimization, where $ \chi^2\equiv \sum_{i=1}^{N_{\rm data}}(\sigma_{{\rm LOS},i}-\sigma_{\rm LOS}(r_i))^2/\sigma_i^2 $ and $ \sigma_i $ is the reported uncertainty for radial bin $ i $. At a given particle mass $m$, our fit has three free parameters: the degeneracy parameter $\mu_0/T$, the central core density $\rho_0$, and the stellar velocity anisotropy parameter $\beta$, taken to be constant in $r$.

\begin{figure}[htbp!]
	\centering
	\includegraphics[width=0.45\textwidth]{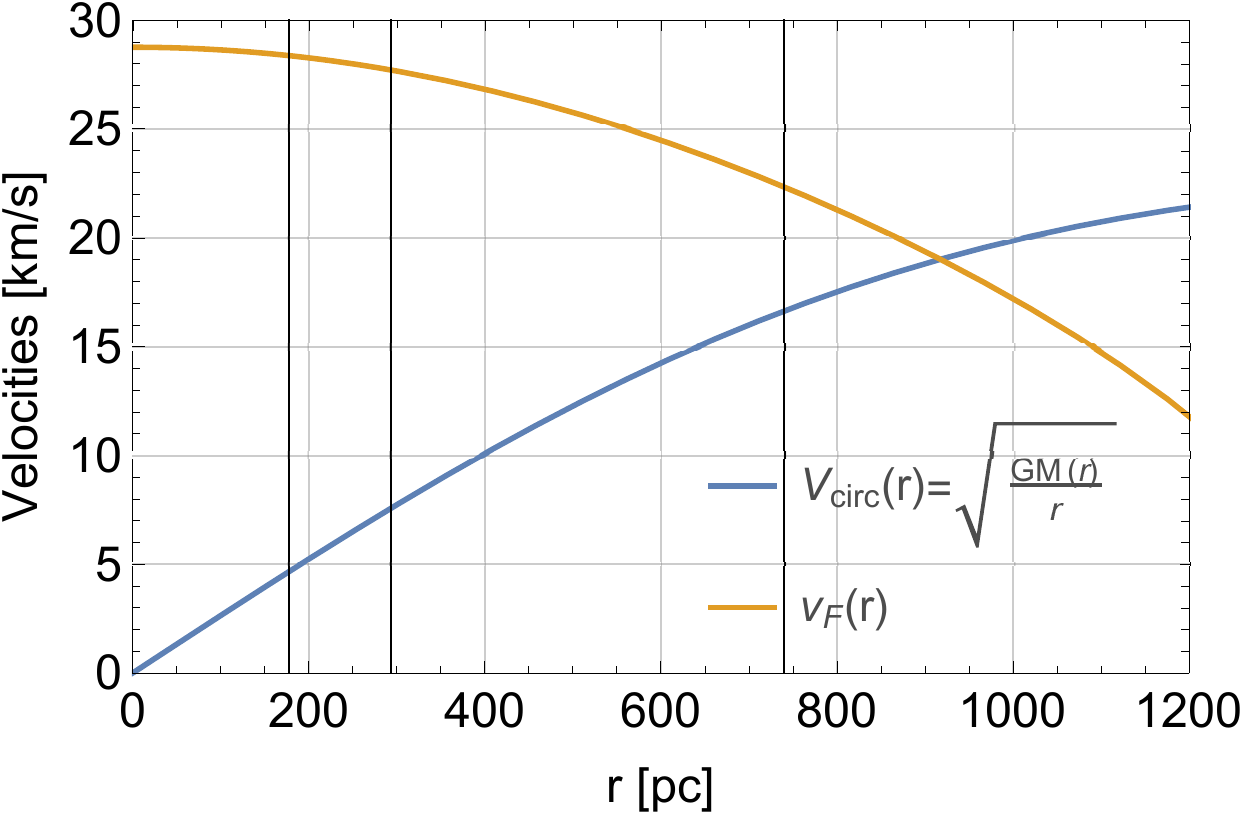}	
	\caption{The circular velocity $ V_{\rm circ} $ and the Fermi velocity $ v_F $ for the DDM halo with $ m=135 $~eV, $ \rho_0=3.9\times 10^{7} M_{\odot}/{\rm kpc}^3 $, $ \mu_0/T=3 $ and $ g=2 $. 
	The vertical lines show the estimated orbital radii ($ r=r_\perp \times 2/\sqrt{3} $) of the three GCs closest to the dynamical center of Fornax, c.f. Table \ref{tab:dat}. }\label{fig:velocity}
\end{figure}

In Fig.~\ref{fig:velocity} we plot the circular velocity compared to the Fermi velocity, which we define using $ v_F(r) \equiv \sqrt{2\mu(r)/m} $, for the $m=135$~eV fit. As can be expected, the circular velocity scales as $ V\propto r $, whereas $ v_F$ remains constant at small radii. This leads to a suppression of DF, as explained in Sec.~\ref{s:cdmcorecusp}. 
%

\subsection{Structure formation constraints on DDM}\label{ss:ddmcosmo}

The free streaming of light DM suppresses the matter power spectrum \cite{Bond:1980ha}, notably constrained via Ly-$\alpha$ forest statistics \cite{Viel:2005qj,Viel:2013apy,Baur:2015jsy}, with details depending on the cosmological DM production mechanism. 
Refs. \cite{Domcke2015,Randall2017} considered non-thermal mechanisms for cosmological production of DDM, aiming to bypass the Ly-$\alpha$ bounds. With such mechanisms in mind, values of $m$ in the ballpark of 100~eV were considered in these works. 
We now revisit the cosmological bound and formulate a conservative limit that is insensitive to the cosmological production mechanism of DM. Our results suggest that mechanisms of the kind proposed in \cite{Domcke2015,Randall2017} should not be able to produce $m<1.4$~keV without tension with the nominal Ly-$\alpha$ bound. 

The instantaneous DM free-streaming wavelength $k_{\rm FS}$ depends on the DM velocity dispersion\footnote{This formula is correct up to an order unity factor relating the speed of sound $c_s$ and the velocity dispersion $\sigma$ \cite{Shoji:2010hm}. This factor is unimportant for our analysis.}, 
\be
k_{\mathrm{FS}}(z) \equiv \sqrt{\frac{3}{2}} \frac{\mathcal{H}(z)}{c_{\mathrm{s}}(z)} \simeq \sqrt{\frac{3}{2}} \frac{\mathcal{H}(z)}{\sigma (z)}.
\ee 
Here $z$ is the redshift, $\mathcal{H}(z)$ is the Hubble rate, we have set $ \sigma(z) = \sqrt{\left<v^2\right>} $ and 
\be
\left<v^2\right> & = & \frac{\int d^3p \frac{p^2}{m^2+p^2}f(p)}{\int d^3p f(p)}.
\ee
Bounds on warm DM (WDM) \cite{Baur:2015jsy} effectively constrain $k_{\mathrm{FS}}(z)$. Specifically, they apply to $z\lesssim10^6$, where density perturbations on comoving scales of the order of $\lambda\approx(1+z)/\mathcal{H}(z)\approx0.5$~Mpc enter the horizon and begin to evolve under their own gravitational potential. 

We can convert the WDM limit of Ref.~\cite{Baur:2015jsy} into a bound on DDM  by the following prescription. 
At a given energy density, the coldest possible distribution function of DDM is the fully degenerate distribution $ f(p) = \theta(p_F-p) $, where the Fermi momentum $ p_F $ is related to the energy density via
\be
\rho & =& \frac{g}{(2\pi)^3}\int d^3p \sqrt{m^2+p^2}f \\& =& \frac{g}{16\pi^2}\left[p_F\sqrt{m^2+p_F^2}(m^2+2p_F^2)-m^2\sinh^{-1}\left(\frac{p_F}{m}\right)\right] \no \\ & \approx & \begin{cases}
	\frac{g mp_F^3}{6\pi ^2} & p_F\ll m \\ 	\frac{g p_F^4}{8\pi ^2} & p_F\gg m
\end{cases} \; .\no
\ee
The $p_F$ parameter redshifts as $p_F\propto(1+z)$.
The velocity dispersion for this distribution is 
\be\label{eq:v2ddm}
\left<v^2\right> & = & 1-3\left(\frac{m}{p_F}\right)^2+3\left(\frac{m}{p_F}\right)^3\arctan\left(\frac{p_F}{m}\right) \no \\ & \approx & \begin{cases}
	\frac{3p_F^2}{5m^2} & p_F\ll m \\ 1 & p_F\gg m
\end{cases} \; .
\ee

\begin{figure}[htbp!]
	\centering
	\includegraphics[width=0.45\textwidth]{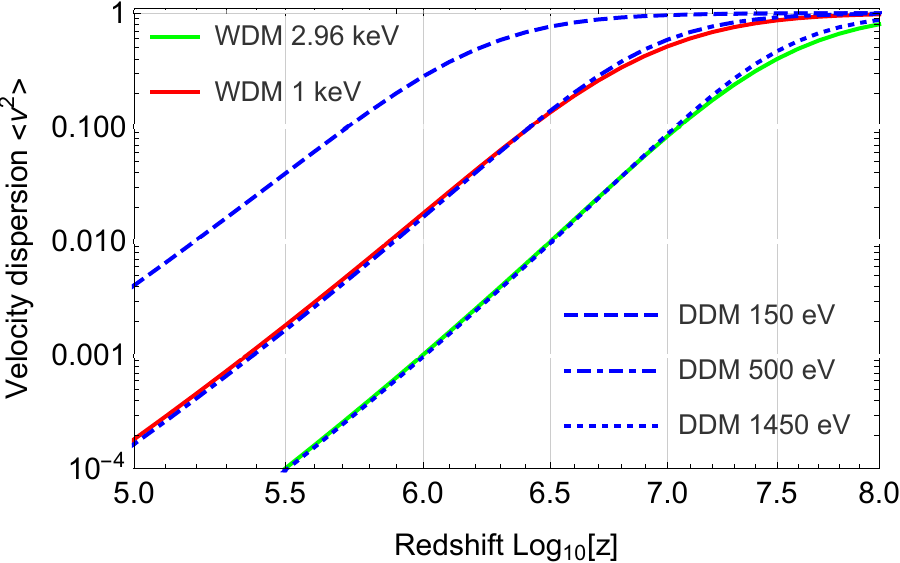}				\caption{A comparison of the velocity dispersions (in natural units) of WDM and DDM for various particle masses. Including CMB data, Ref.~\cite{Baur:2015jsy} puts a bound of $ m>2.96 $~keV on WDM ($ 95\% $ C.L.), for which we plot the velocity dispersion as the green solid line.}\label{fig:lymanalpha}
\end{figure}

Using Eq.~(\ref{eq:v2ddm}) we can calculate $k_{\mathrm{FS}}(z)$, compare this to the $k_{\rm FS}(z)$ of WDM, and cast the bounds of Ref.~\cite{Baur:2015jsy} into the most conservative, maximally-cold DDM model by matching the $k_{\rm FS}(z)$ curves of the two models. 
To recall, WDM was defined \cite{Baur:2015jsy} by the distribution function $ f_{\rm WDM} = \left(\exp(p/T)+1\right)^{-1} $, where $T\propto(1+z)$. 
Comparing the DM mass density for DDM and WDM in the nonrelativistic regime, we have
\be
\rho &=& \begin{cases}
	\frac{gm p_F^3}{6\pi^2} & {\rm DDM} \\ \frac{3\zeta(3)g mT^3}{4\pi^2} & {\rm WDM} 
\end{cases} \; .
\ee
Matching the density in the two models implies:
\be\label{eq:rhomatch} \frac{\left(T/m\right)_{\rm WDM}}{\left(p_F/m\right)_{\rm DDM}}&=&\left[\frac{2}{9\zeta(3)}\frac{\left(gm^4\right)_{\rm DDM}}{\left(gm^4\right)_{\rm WDM}}\right]^{\frac{1}{3}}.\ee
On the other hand, still in the nonrelativistic regime we can compare the velocity dispersion in the two models,
\be\frac{\langle v^2\rangle_{\rm DDM}}{\langle v^2\rangle_{\rm WDM}}&\approx&\frac{3\zeta(3)}{75\zeta(5)}\left[\frac{\left(p_F/m\right)_{\rm DDM}}{\left(T/m\right)_{\rm WDM}}\right]^2\no\\
&=&\frac{3\zeta(3)}{75\zeta(5)}\left(\frac{9\zeta(3)}{2}\right)^{\frac{2}{3}}\left[\frac{\left(gm^4\right)_{\rm WDM}}{\left(gm^4\right)_{\rm DDM}}\right]^{\frac{2}{3}},\ee
where in the second line we used Eq.~(\ref{eq:rhomatch}). 
We illustrate the comparison in Fig.~\ref{fig:lymanalpha}. For model parameters of interest to this discussion, the ratio of velocity dispersions is redshift-independent at $z\lesssim10^6$. We can therefore conclude that the WDM constraint of Ref.~\cite{Baur:2015jsy}, $m_{WDM}>2.96$~keV at the 95\%C.L. for $g=2$, implies the constraint:
\be\label{eq:ddmcosmobound} gm^4_{DDM}&>&2\times\left(1.4~{\rm keV}\right)^4.\ee

We expect that model-building around the bound of Eq.~(\ref{eq:ddmcosmobound}) would be quite difficult. No production mechanism should be able to create a colder distribution function for DDM. In particular, the skewed momentum distribution scenarios of Ref.~\cite{Randall2017} and the scalar decay models of Refs.~\cite{Domcke2015,Choi:2020nan} should all satisfy this bound. Using Eq.~(\ref{eq:rcddm}), we find that Eq.~(\ref{eq:ddmcosmobound}) constrains the DDM core in Fornax to $r_c\lesssim20$~pc, irrelevant for the orbits of GCs. 

The Ly-$\alpha$ analyses may be affected by systematic uncertainties related, among other things, to the thermal history of the inter-cluster medium and other baryonic effects \cite{Baur:2015jsy}. Keeping this caveat in mind, it seems sensible to take Fig.~(\ref{fig:lymanalpha}) with a grain of salt. If we allow $k_{\rm FS}$ of the (coldest possible) DDM model to exceed the nominal bound of Ref.~\cite{Baur:2015jsy} by, say, a factor of $\sim4.5$, we could relax Eq.~(\ref{fig:lymanalpha}) to $gm^4\,>\,2\times\left(500~{\rm eV}\right)^4$. With such a (rather ad-hoc) relaxed bound we could allow a DDM core $r_c\lesssim80$~pc, still irrelevant for GC orbits in Fornax. Going down to $m=150$~eV (still at $g=2$), which would allow a DDM core radius of $r_c\approx385$~pc encompassing some GC orbits, would amount to $k_{\rm FS}$ being smaller by a factor of 20 than the nominal WDM bound.

\subsection{Orbital decay time in DDM}\label{sec:decay}
As we have seen, Ly-$\alpha$ analyses exclude DDM from producing a core extending to the observed orbital positions of GCs in Fornax. This means that DDM would not change the standard CDM predictions for the DF settling time of the GCs. Nevertheless, given that the Ly-$\alpha$ bound is subject to some debate, it is interesting to see what DDM could do to DF subject only to the LOSVD constraints of Sec.~\ref{ss:ddmlosvd}.

In order to estimate the instantaneous DF time scale $ \tau $, we use \refeq{genDecaytime2}, with a modified \refeq{Cdeg}. In order to interpolate between a quasi-degenerate core and a classical gas in the outskirts of the halo, we adopt
\be
C_{\rm DDM} = \frac{1}{\frac{1}{0.5}+\frac{v_F^3}{V^3}}\ln\Lambda \; ,
\ee
such that in the regime $ V\gg v_F $, we retrieve CDM-like behavior, c.f. Sec.~\ref{s:cdmcorecusp}. For $ \ln\Lambda $, we adopt the choice for $ \ln\Lambda_{\rm ISO} $, c.f. \refeq{clogISO}.

We use the GC masses and projected radii collected in Table~\ref{tab:dat}, combined with the density profile derived in the LOSVD fits. We  correct for the projection effect by relating the assumed true orbital radius to the observed projected radius of the GC via $ r_{\rm true}/r_{\perp}= 2/\sqrt{3} $. We also assume that the GCs are on circular orbits, setting $ V_{\rm true}/V_{\rm circ}(r_{\rm true})=1 $. (This de-projection procedure is, of course, simplistic: we will shortly report a more comprehensive treatment.) The results are summarized in Table~\ref{tab:dat}. 
For $ m=135 $~eV we find na\"ive orbital decay time-scales of $ 1.48 ~$Gyr and $ 10.7 $~Gyr for GC3 and GC4, respectively. 
For comparison, using the approach of Ref. \cite{Hui2017} for cuspy CDM we find $ 2.63 $~Gyr and $ 0.99 $~Gyr. Therefore, while the na\"ive DF time of GC4 in DDM is much longer than in cuspy CDM, for GC3 the na\"ive time in cuspy CDM actually exceeds that of DDM. 

However, as discussed in Sec.~\ref{s:cdmcorecusp}, the instantaneous $\tau$ can be misleading when comparing different halo morphologies: a realistic estimate of the GC settling time requires orbit integration. We turn to this analysis next, finding that the real orbit settling times of both GC4 and GC3 are in fact  longer in DDM compared with cuspy CDM.

To obtain a more comprehensive estimate of the DF settling time and the impact of projection effects, we use the orbit integration explained in App.~\ref{app:orbits} with initial conditions that we vary as follows. For each GC, we scan the range $ r_{\rm true}\in [1,2] r_{\perp} $ (the logic behind this range is explained in App.~\ref{app:eccentricity}). For each $ r_{\rm true} $ we scan over $ V_{\rm true} \in [0.5,1.5]V_{\rm circ}(r_{\rm true})$. For each $V_{\rm true}$ we test positive and negative $ \cos\theta $. Finally, we test the two cases, $ \Delta v_y= \sqrt{v_{\rm true}^2-\Delta v_r^2}, \Delta v_z=0 $ and $ \Delta v_y=0, \Delta v_z=\sqrt{v_{\rm true}^2-\Delta v_r^2} $. 

For each starting point in phase-space, we integrate the equations of motion, stopping the integration when $ (r_{\rm apoenter}+r_{\rm pericenter})/2 \lesssim 0.3 r_{\rm initial}\equiv f_r r_{\rm initial} $, or after $ 10$~Gyr (the earlier of the two). We then report the integration time as $ \tau_{\rm inspiral} $.

\begin{figure}[htbp!]
	\centering
	\includegraphics[width=0.45\textwidth]{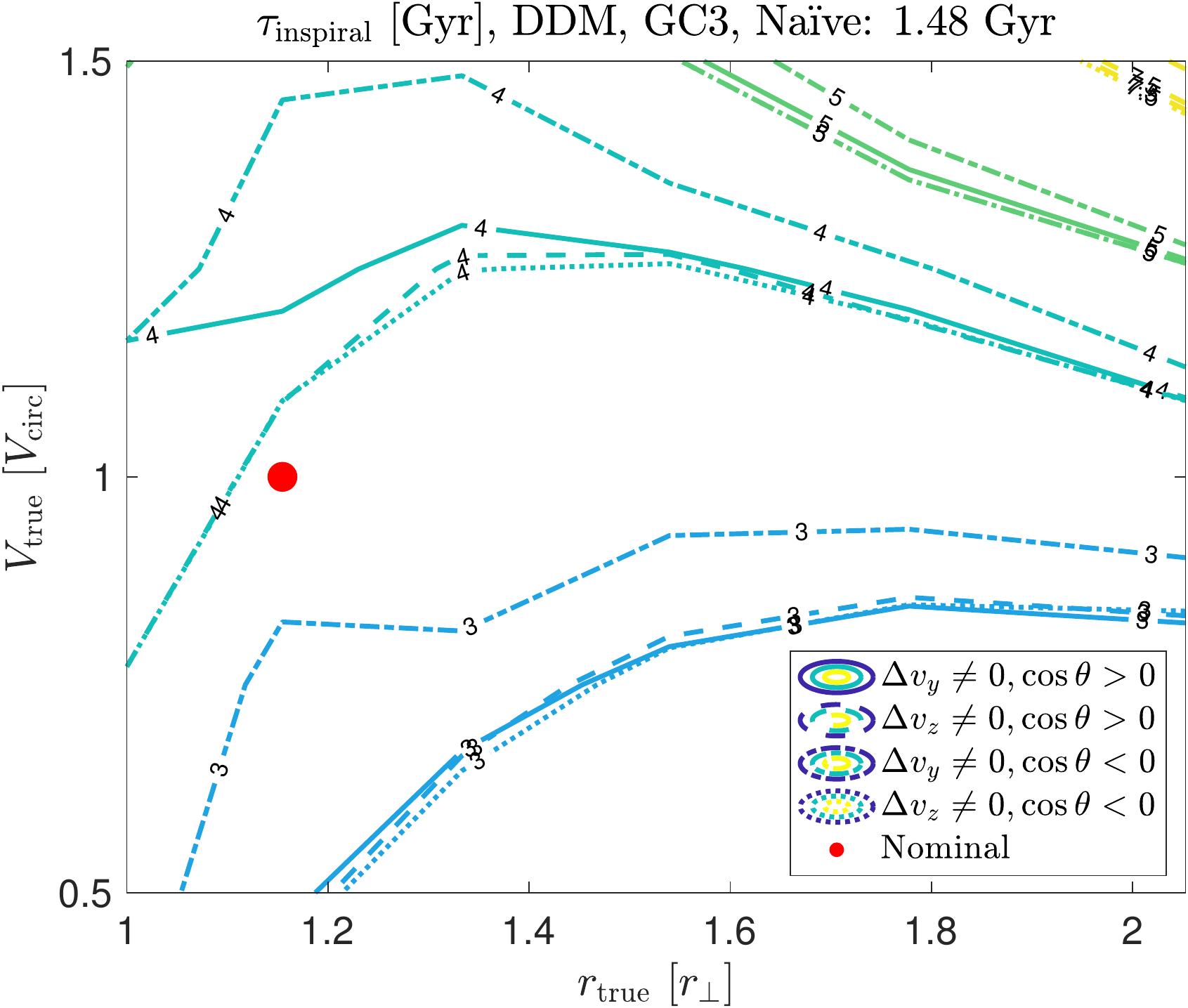}\\	\includegraphics[width=0.45\textwidth]{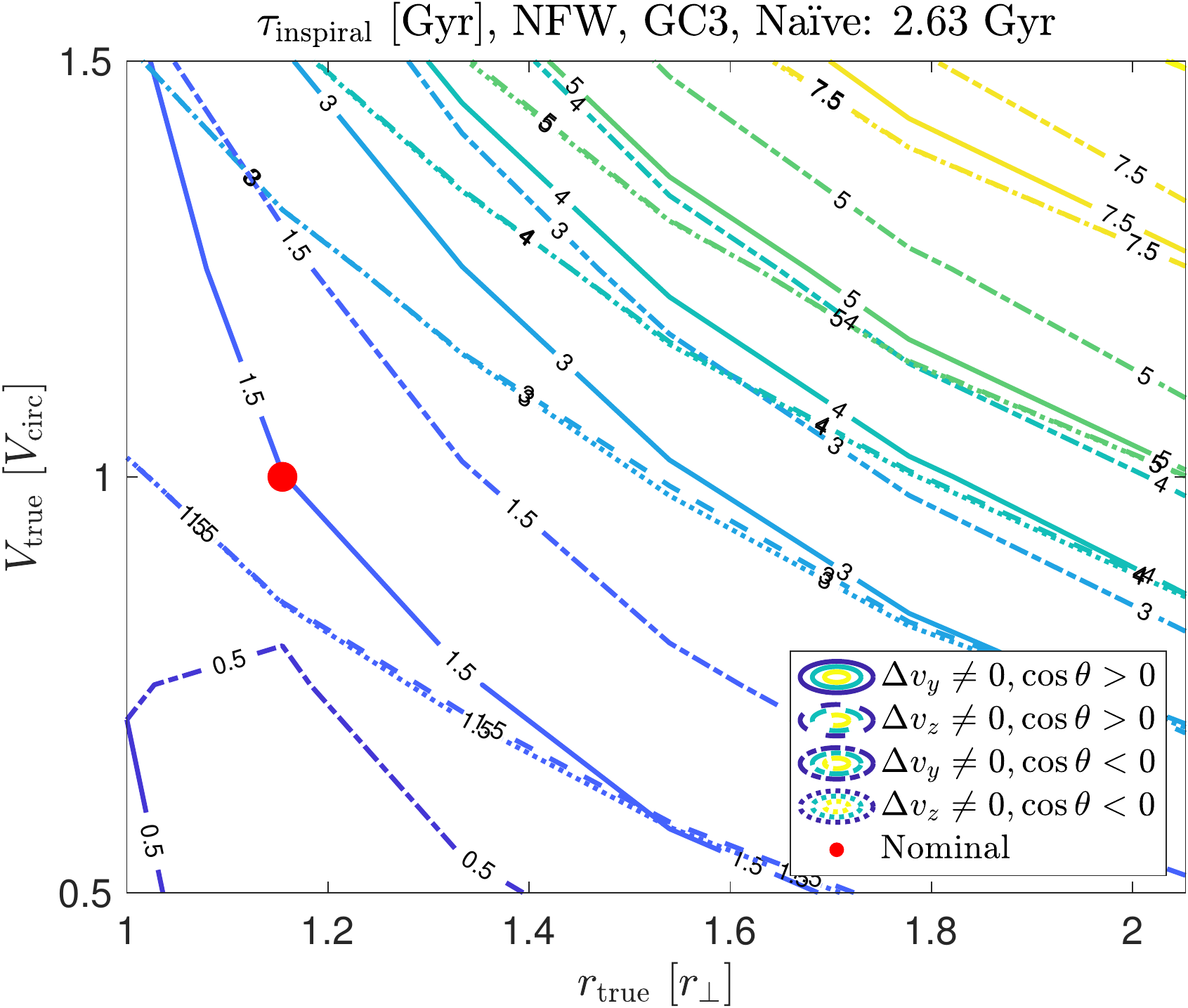}
	\caption{Contours of the inspiral time of GC3, defined in App.~\ref{app:eccentricity}, for the DDM (\textbf{top}) and the cuspy CDM models (\textbf{bottom}). The ``na\"ive'' estimates written on top are those given in Table~\ref{tab:dat} based on an evaluation of the instantaneous DF time at $ r_{\rm true}/r_{\perp}=2/\sqrt{3} $ and $ V_{\rm true}/V_{\rm circ}(r_{\rm true})=1 $.  
	The different line types correspond to different discrete choices in our scan of the initial conditions in phase-space, explained in more detail in App.~\ref{app:eccentricity}.}\label{fig:GC3cmp}
\end{figure}

In Fig.~\ref{fig:GC3cmp} we plot the result of this procedure for GC3, comparing the DDM halo for $ m=135 $~eV (top panel) and the cuspy CDM halo from Ref.~\cite{Meadows20} (bottom panel). For the representative phase-space point $ r_{\rm true}/r_{\perp}=2/\sqrt{3} $, $ V_{\rm true}/V_{\rm circ}(r_{\rm true})=1 $, highlighted in Fig.~\ref{fig:GC3cmp} by a red dot, we find that the inspiral time in the DDM halo is in fact longer ($ \sim 4 $~Gyr) than in the cuspy CDM one ($ \sim 1.5 $~Gyr). This result is in reversed order to the nai\"ve estimate in Table~\ref{tab:dat}, demonstrating that the nai\"ve DF time estimate can indeed be misleading.

We note that the inspiral time in the NFW case is not very sensitive to the stopping radius fraction $ f_r $ (set as $0.3 $), whereas the DDM case is, and so are other cored halo models. As explained in Sec.~\ref{s:cdmcorecusp}, a cuspy profile predicts approximately $ \tau(r)\propto r^{2} $, therefore the inspiral time is mostly sensitive to the initial radius. In a cored model, $ \tau(r) $ is a weak function of $ r $, potentially even non-monotonuous. Therefore, the definition of the inspiral time in the cored model becomes sensitive to the radius at which the orbit integration is stopped.

\section{Self-interacting dark matter (SIDM)}\label{sec:sidm}
Self-interacting DM (SIDM) is a simple modification of CDM, that could arise in many models \cite{Spergel:1999mh,Kaplinghat:2013xca,Kaplinghat:2015aga,Rocha:2012jg,Sokolenko:2018noz,Zavala_2013,Tulin:2017ara,Fitts:2018ycl,Robertson:2020pxj}. The self interactions can be expressed in terms of the cross section per unit mass, $ \sigma/m$
, which could be velocity-dependent \cite{Tulin:2017ara}. The scattering mean free path is
\be\label{eq:mfpSIDM}
l &=& \frac{m}{\rho \sigma}\; =\; 48\frac{10^8~M_{\odot}/{\rm kpc}^3}{\rho}\frac{1~{\rm cm}^2/{\rm gr}}{\sigma/m}~{\rm kpc}  ,
\ee
and the time between scatterings $ l/v $ is
\be
t_{\rm scat} &=& 2.35\frac{20~{\rm km/s}}{v}\frac{10^8~M_{\odot}/{\rm kpc}^3}{\rho}\frac{1~{\rm cm}^2/{\rm gr}}{\sigma/m}~{\rm Gyr} . 
\ee

When $ l $ is larger than the distance across the halo, we expect that the microphysics of DF in the SIDM model will be similar to that of CDM. On the other hand, the morphology of an SIDM halo could be different to that in CDM as long as $t_{\rm scat}$ is smaller than the age of the system. Given a large enough cross section, SIDM produces cored halos which affect the orbital settling time of GCs as discussed in Sec.~\ref{s:cdmcorecusp}. 

We follow Ref.~\cite{Kaplinghat:2015aga} in modeling the SIDM halo profile. Inside some radius $ r_1 $, we assume a hydrostatic profile with central density $ \rho_c $ and pressure $ P = \sigma_0^2\rho $. For a self-gravitating spherical halo, the density profile obeys
\be
\frac{1}{r^2}\partial_r (r^2\partial_r\ln\rho_{\rm iso}) = -\frac{1}{r_{\rm ic}^2}\frac{\rho_{\rm iso}}{\rho_c} \; ,
\ee
where $ r_{\rm ic} \equiv  \sigma_0/\sqrt{4\pi G\rho_c} $ (similar to the King radius \cite{BinneyTremaine2}). Beyond $ r_1 $ we match the density to the NFW profile, $\rho_{\rm NFW}= \rho_s(r/r_s)^{-1}(1+r/r_s)^{-2} $, fixing $ \rho_s $ and $ r_s $ by imposing continuity of $ \rho_{\rm iso}(r_1) = \rho_{\rm NFW}(r_1) $ and of the enclosed mass $ M_{\rm iso}(r_1) = M_{\rm NFW}(r_1) $. This procedure is consistent with an initially NFW-like cusp profile that was deformed into a cored isothermal profile due to the SIDM scatterings. Altogether, the halo model has three free parameters, $ \rho_ c$, $ r_1 $ and $ \sigma_0 $, that we can constrain with LOSVD data.

\begin{figure}[htbp!]
	\centering
	\includegraphics[width=0.45\textwidth]{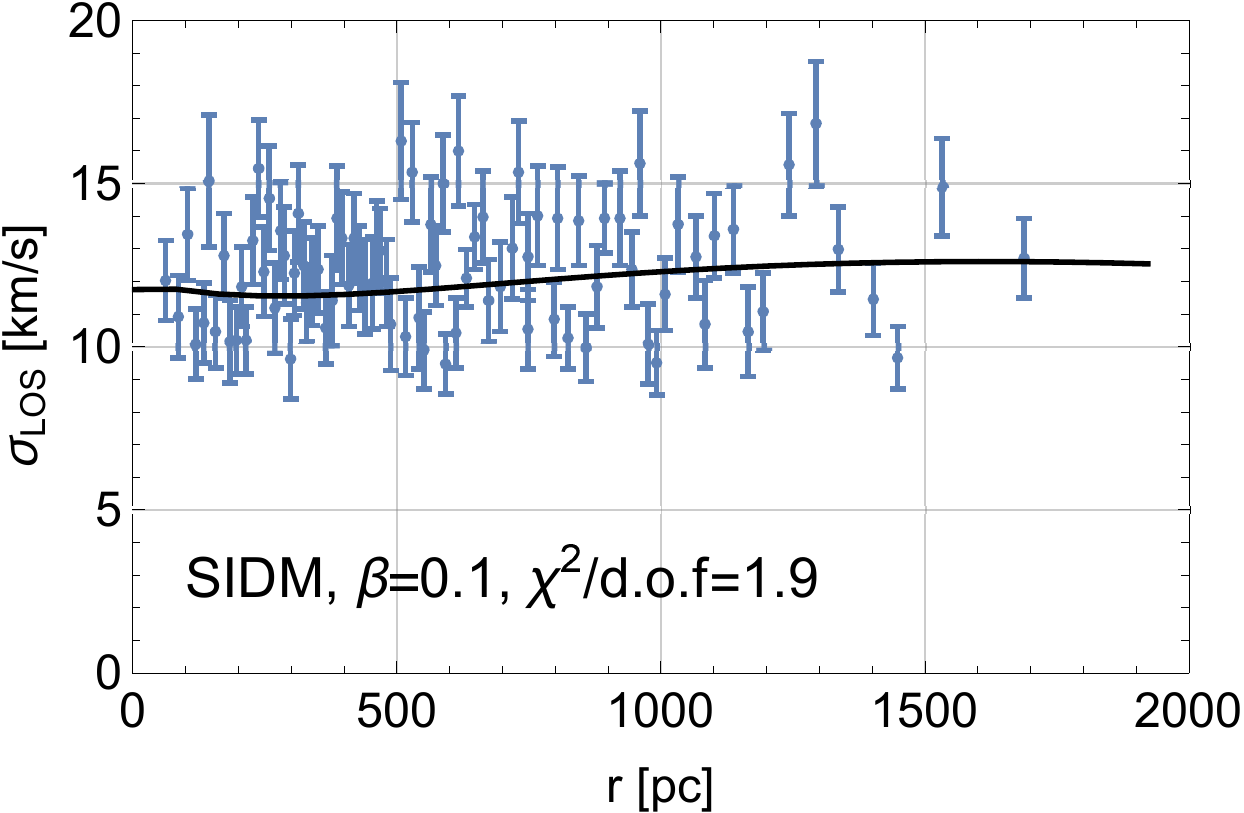}
	\caption{LOSVD data of Fornax dSph modeled by different SIDM profiles. The central density is $ \rho_c=2.6\times 10^{7}~M_{\odot}/{\rm kpc}^3 $, the velocity dispersion is $ \sigma_0=17~{\rm km/sec} $ and $ r_1/r_{\rm ic}=6 $.}\label{fig:losvdSIDM}
\end{figure}

In \reffig{losvdSIDM} we plot a LOSVD fit, following the same $ \chi^2 $ procedure as in Sec.~\ref{sec:ddm}. The model is taken to illustrate a large core solution that would include the orbit of GC3.

\begin{figure}[htbp!]
	\centering
	\includegraphics[width=0.45\textwidth]{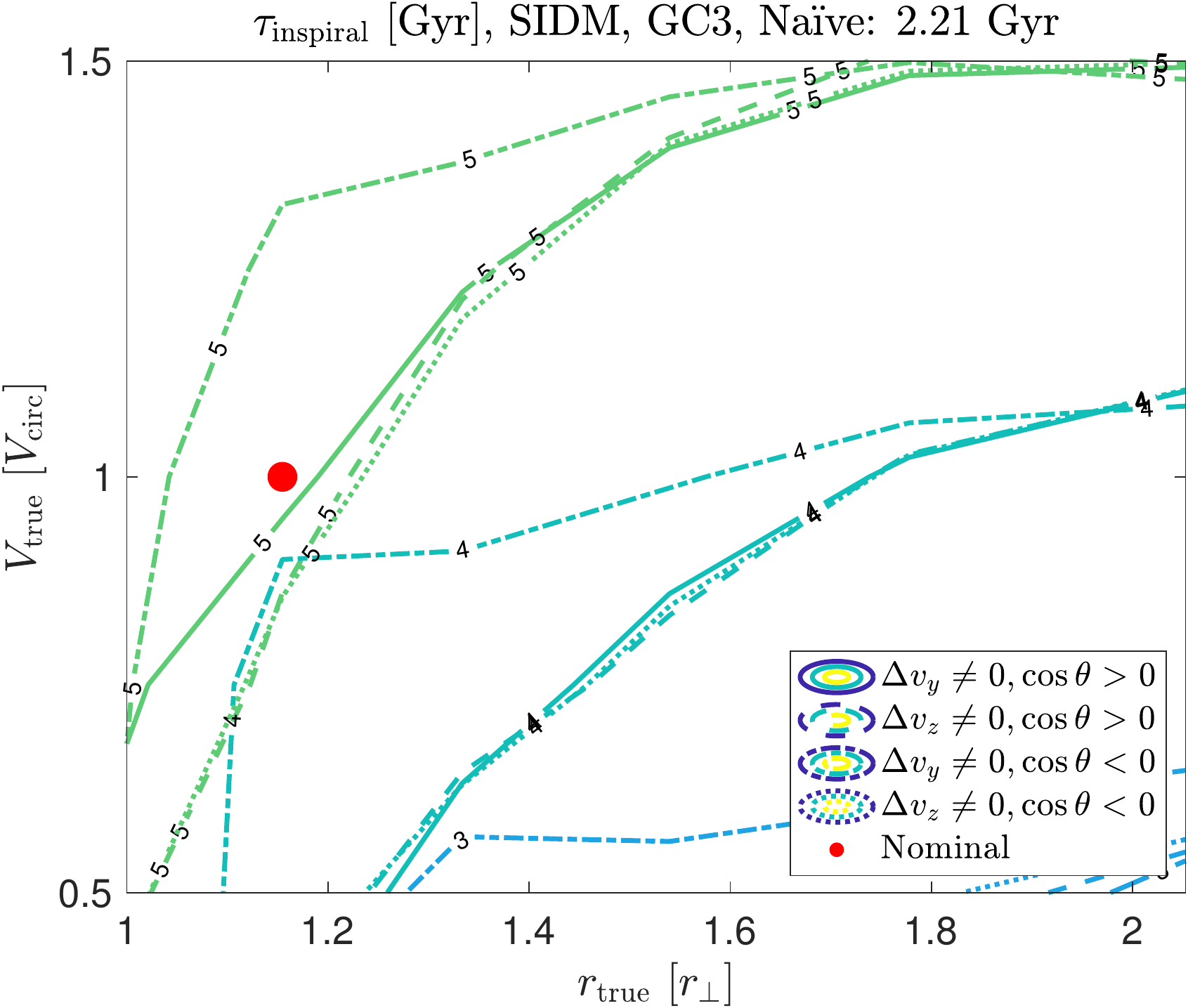}	
	\caption{DF time for the SIDM halo, analogously to \reffig{GC3cmp}.}\label{fig:inspiralSIDM}
\end{figure}

In \reffig{inspiralSIDM} we show the inspiral time of GC3, using the same procedure as in \reffig{GC3cmp}. We use \refeq{CMax} with $ \sigma_0$ from the LOSVD fit and adopt $ \ln\Lambda=\ln\Lambda_{\rm ISO} $ as in \refeq{clogISO}. We find that the large core SIDM model significantly increases the inspiral time of GC3 compared to the CDM prediction. 

Ref. \cite{Kaplinghat:2015aga} pointed out that in the SIDM core region, DM particles have undergone about a single collision during the age of the system, i.e.
\be
\frac{\left<\sigma v\right>}{m}\rho_c t_{\rm age}&=& \frac{t_{\rm age}}{t_{\rm scat}}\;\approx\;1 \; .
\ee
With this assumption,\footnote{It has been pointed out \cite{Bondarenko:2020mpf,Sokolenko:2018noz} that this assumption may be  simplistic. } we can estimate the cross section implied by the LOSVD fit
\be
\frac{\left<\sigma v\right>}{m} &\approx & 18\frac{10~{\rm Gyr}}{t_{\rm age}} \frac{2.6\times 10^{7}M_{\odot}/{\rm kpc}^3}{\rho_c} {\rm\frac{cm^2}{g}\frac{km}{s}} \; .
\ee
This result is compatible with the baseline model of Ref.~\cite{Kaplinghat:2015aga}, which predicted $\left<\sigma v\right>/m \sim 25~{\rm cm}^2{\rm g}^{-1}{\rm km~ s}^{-1} $. It would significantly increase the DF settling time of the innermost Fornax GCs.
%

\section{Discussion}\label{sec:discussion}

\begin{figure*}[htbp!]
	\centering
	\includegraphics[trim={0cm 0 0 0cm}, width=0.46\textwidth]{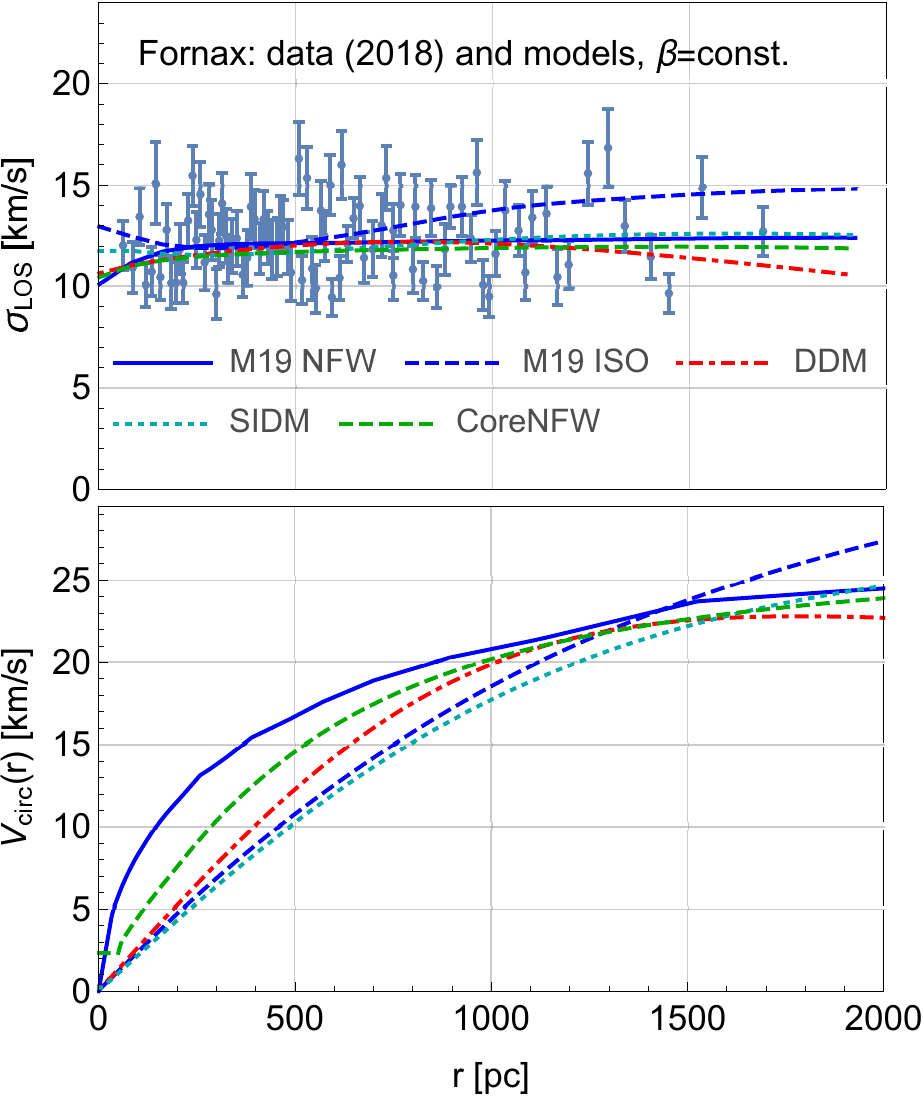}	\includegraphics[trim={0cm 0cm 0 0.0cm}, width=0.474\textwidth]{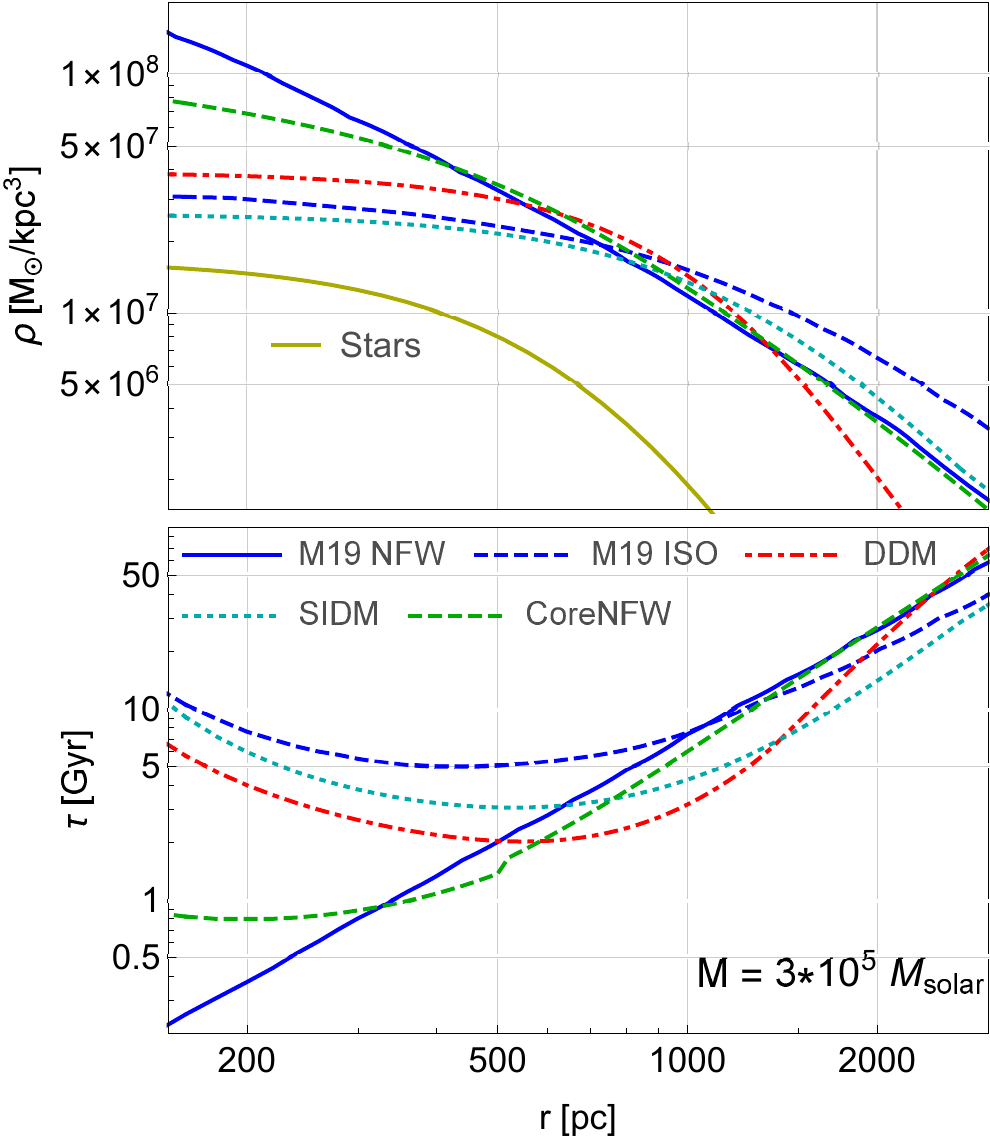}
	\caption{Comparison of models. {\bf Top left:} LOSVD data and fits. The M19 NFW and M19 ISO models refer to the halos of \cite{Meadows20}, for which we only fit the velocity anisotropy. The DDM and SIDM models are based on Sec.~\ref{sec:ddm} and Sec.~\ref{sec:sidm}. The $ \chi^2 $/d.o.f is $ \approx 1.9 $ for NFW, DDM and SIDM, and $ \approx 2.5 $ for ISO. The velocity anisotropy is taken to be constant in each fit. We find $ \beta_{\rm NFW}=-0.4 $, $ \beta_{\rm ISO}=0.2 $, $ \beta_{\rm DDM}=-0.1 $, $ \beta_{\rm SIDM}=0.1 $ and $ \beta_{\rm coreNFW}=-0.1 $. 
	{\bf Top right:} Density profiles. In addition to DM, we also plot an estimate of the stellar density, assuming a Plummer profile with scale $ r_p=851 $~pc \cite{wang2019morphology} and mass $ 4\times 10^7~M_{\odot} $ \cite{de2016four}. %
	{\bf Bottom left:} Circular velocity. (Note how cored models require some tuning to explain the large radial velocity of GC4, $ |\Delta v_r|= 8.26\pm 0.64 $~km/s, at its small projected radius $ r_\perp \approx 0.154 $~pc). 
	{\bf Bottom right:} Instantaneous DF time, evaluated for $ \M=3\times 10^{5}~M_{\odot} $.} \label{fig:comparemodels}
\end{figure*}

In \reffig{comparemodels} we summarize the key features of different models of DM discussed in this work, including both cuspy and core halo models. In the spirit of Ref.~\cite{Cole2012}, we also add the hybrid  \textsf{coreNFW} model of \cite{Read:2015sta} with density\footnote{For the \textsf{coreNFW} model we adopt the best-fit of Ref.~\cite{Read:2018fxs}, with $ \rho_0 \approx 10^{7.1}~M_{\odot} $, $ r_s\approx 2.1 $~kpc, $ n\approx 0.8 $ and $ r_c\approx 0.52 $~kpc. We derive the velocity dispersion of the halo by solving the Jeans equation, \refeq{veljeans}, assuming isotropy.}
\be
\rho_{\rm coreNFW}=\widetilde{f}^n\rho_{\rm NFW}+\frac{n \widetilde{f}^{n-1}(1-\widetilde{f}^2)}{4\pi r^2r_c}M_{\rm NFW} \; , \label{Baryon_Cusp}
\ee
where $ \widetilde{f} = \tanh(r/r_c) $ and $ M_{\rm NFW} = \int_0^rd^3r^{\prime}\rho_{\rm NFW} $. This model aims to describe a CDM-dominated halo modified by baryonic feedback.

The LOSVD data (top left panel of \reffig{comparemodels}) is described reasonably well in all models, with the fit of the ISO model of Ref.~\cite{Meadows20} being slightly worse. 

The instantaneous DF time for a GC with $m_*=3\times10^5$~M$_\odot$ is shown in the bottom right panel of \reffig{comparemodels}. It illustrates the fact that the main impact of the microphysics of DM (as in DDM and SIDM) on DF comes from their prediction of a cored halo morphology, and not from the exotic microphysics per-se. The formation of a core due to baryonic feedback in CDM \cite{Goerdt2006,Read:2006fq} could therefore have similar consequences. 

The density profiles (top right panel of \reffig{comparemodels}) demonstrate the cusp for NFW, large cores and the intermediate \textsf{coreNFW}. We also plot an estimate of the stellar density. This may become important for large-core models, whose density is only larger by a factor of $ 2 $ or so than the stellar density at small $r$. 
In these cases, accounting for the stellar-induced potential could slightly change our results numerically, but not qualitatively: as far as the DF microphysics is concerned, background stars would contribute to the DF of a GC just like CDM particles, and since the total mass density is constrained by the LOSVD fit, the separation into DM and stellar components is not essential for the DF computation. 

It is interesting to compare the circular velocity profiles of different models (bottom left panel of \reffig{comparemodels}) to the phase-space parameters of GCs. 

In Sec.~\ref{ss:gcbygc} we briefly consider each of the six GCs, noting the implications w.r.t. the timing puzzle. Next, in Sec.~\ref{ss:stat} we consider the GC system as a whole and use the tools we have developed throughout this work to re-evaluate the problem.

\subsection{Comments regarding GC data used in this work (GC-by-GC discussion)}\label{ss:gcbygc}
Let us re-evaluate the role of each GC in the timing puzzle, in light of the observational data used in this work (see Tab.~\ref{tab:dat}) including projected radii that are different than those of earlier analyses \cite{Mackey2003a,Cole2012,Hui2017,boldrini2020embedding} and LOS velocities that were often ignored in past analyses, but are in fact known fairly well from observations. 
\begin{enumerate}
	\item GC1 and GC5 -- these GCs do not seem to pose a timing problem as they are located at fairly large projected radii, $ r_{\perp}\approx  1.7 $~kpc, not far below the (somewhat model-dependent) tidal radius of Fornax at $ 1.8\div 2.8 $~kpc \cite{angus2009resolving,Cole2012,Read:2005zm}. 
	That said, it is interesting to note that while the circular velocity at the GC radii is $ 20\div 30 $~km/s, the measured GC LOS velocities are smaller than $V_{\rm circ}$ by a factor of 5 or so. This could hint that GC1 and GC5 are close to the apocenter of fairly eccentric orbits. If this is indeed the case, then the na\"ive instantaneous DF time overestimates the true orbital settling time because the GCs typically experience stronger DF when they venture into smaller radii, as expected if the orbit is eccentric. 
	
	\item GC2 -- does not seem to present a timing problem.
	
	\item GC3 -- for a cuspy profile, our ``na\"ive'' instantaneous DF time of $ 2.6 $~Gyr comes in some disagreement with the $ \tau \approx 0.6 $~Gyr quoted in Ref.~\cite{Hui2017}. The main reasons for the difference are the new estimate of $ r_\perp $ and the updated LOSVD data \cite{Read:2018fxs} compared to the older data \cite{walker2009universal}. Of more relevance, however, is the actual physical inspiral time obtained with the orbit integration of App.~\ref{app:eccentricity}. We find that both a simple cusp profile and the intermediate \textsf{coreNFW} profile predict an inspiral time $ \approx 1.5 $~Gyr, whereas models with a large core predict $ 4\div 5 $~Gyr. 
	
	\item GC4 -- the cuspy CDM orbital decay time is short $ \approx 1 $~Gyr, but not as short as previously estimated \cite{Hui2017}. Beyond the reasons listed for GC3, we find that the approximation $ C\approx 0.5\ln\Lambda $, adopted in Ref.~\cite{Hui2017}, is not accurate for small radii, c.f. Sec.~\ref{s:cdmcorecusp}.
	
	A large core would stabilize the orbit of this GC to the 10~Gyr time scale, and even 
	the intermediate \textsf{coreNFW} profile predicts an inspiral time of $ \approx 5$~Gyr.
	
	GC4 is younger and more metal-rich compared to the other GCs \cite{Mackey2003a,de2016four}, and it has been debated in the literature whether it is in fact the nuclear star cluster of Fornax \cite{Hardy2002,Strader_2003,de2016four,Martocchia_2020}. The LOS velocity of this GC, $ \gtrsim 8 $~km/s, appears to potentially be at odds with this interpretation \cite{Hendricks_2015}. 
	
	The large LOS velocity is also somewhat difficult to accommodate within a large core halo model. In all of our cored halo models (see bottom left panel of \reffig{comparemodels} and note that $r_\perp\approx150$~pc for this GC), GC4 needs to be on a circular orbit with $ r_{\rm true}/r_\perp\gtrsim 2 $ or close to the pericenter of an eccentric orbit with $ r_{\rm true}/r_\perp\gtrsim 1.6 $. 
	These possibilities are somewhat tuned, either w.r.t. the projection angle or w.r.t. the orbital phase. In comparison, an NFW profile can comfortably accommodate the radial velocity of GC4. For the \textsf{coreNFW} model, it is marginally possible to have GC4 on a circular orbit without tuning in radius. 
	
	\item GC6 -- the newly rediscovered GC \cite{wang2019rediscovery} probably has a smaller mass ($ \approx 0.29\times 10^{5} ~M_{\odot}$ \cite{Shao:2020tsl}) than the other five GCs. It does not seem to reinforce the timing puzzle. 
	
	Ref.~\cite{wang2019rediscovery} noted that GC6 has an elongated shape and may be undergoing tidal disruption. This may comprise some evidence in favor of a cuspy halo. 

\end{enumerate}

To summarize, the usual suspects for a GC timing puzzle, GC3 and GC4, are found here to have somewhat longer settling times than previously thought \cite{Hui2017}, but nevertheless much shorter than their age. 

\subsection{Statistical discussion}\label{ss:stat}

Let us finally use the tools we developed to re-evaluate the timing puzzle. Consider the cuspy NFW and the cored ISO profiles in Fig.~\ref{fig:comparemodels}. 
Using the results developed in~\refapp{cdfgcs}, we can map a distribution of GC initial radii into the cumulative distribution function (CDF) of GC projected radii today. 

\begin{figure*}[htbp!]
	\centering
	\includegraphics[width=0.5\textwidth]{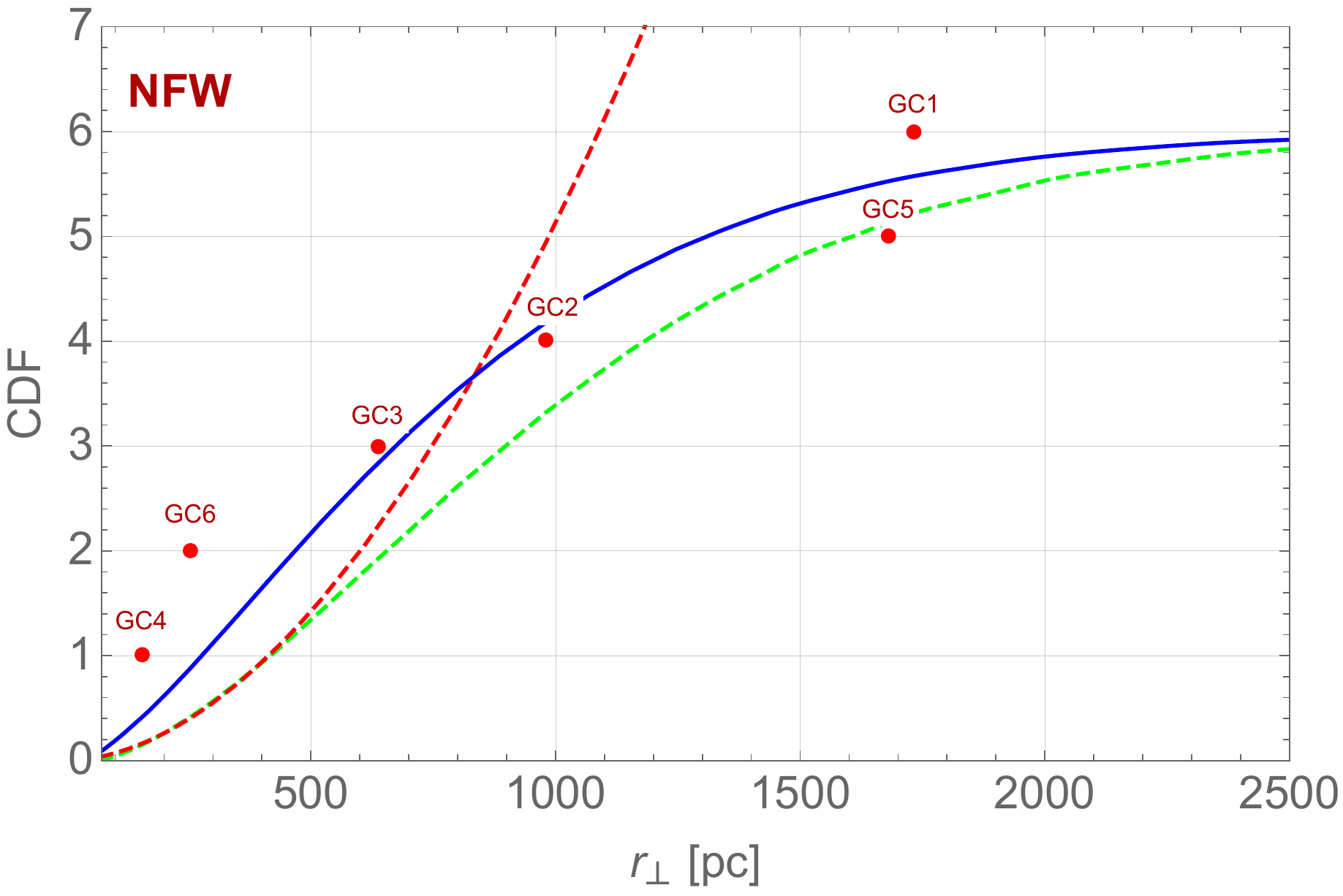}
	\includegraphics[width=0.5\textwidth]{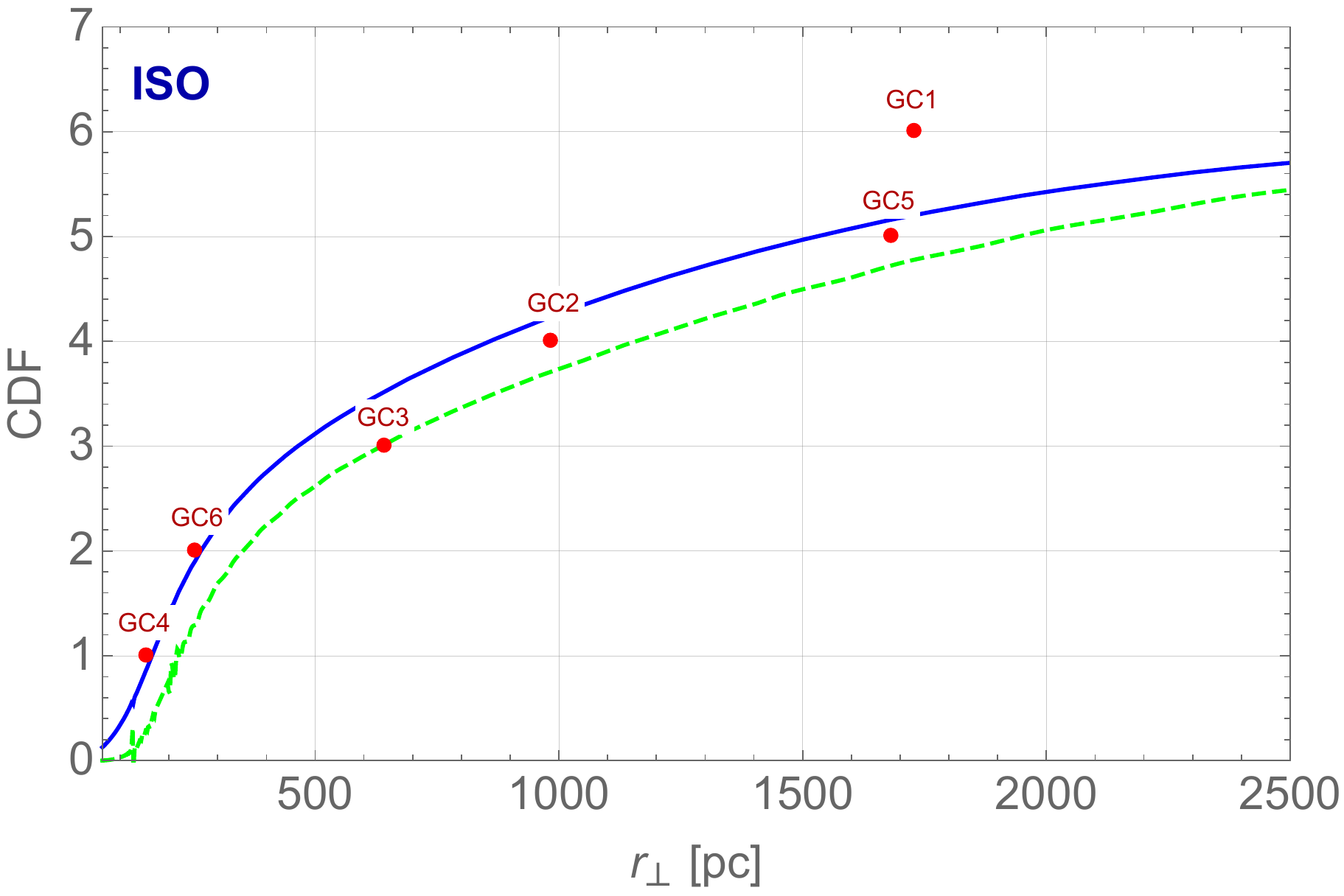}	
	\caption{Example of a calculated CDF of projected radii using $\tau(r)$ from Fig.~\ref{fig:comparemodels}. {\bf Left:} NFW halo. {\bf Right:} Cored ISO halo. The solid blue line shows the CDF after projection effects are taken into account. The dashed green line shows the result before projection. For the NFW case, the small-$r$ prediction of Eq.~(\ref{eq:FDtsum}) is shown by the red dashed line. Observed Fornax GCs are also shown. The initial radial GC PDFs used to make the plot are explained in the text.}\label{fig:FDtperp}
\end{figure*}

An example of such a calculation is shown in Fig.~\ref{fig:FDtperp}, with the NFW result in the left panel and the ISO in the right. For concreteness, in making Fig.~\ref{fig:FDtperp} we used an initial distribution of GC radii of the form $f_0(r_0)\propto r_0^2\exp\left(-a_0r_0\right)$, with $r_0$ given in kpc and the parameter $a_0$ in kpc$^{-1}$. We give this initial distribution $\Delta t=10$~Gyr to evolve. For the NFW example we set $a_0^{-1}=0.3$~kpc, while for the cored ISO case we set $a_0^{-1}=0.6$~kpc. We stress that this form for $f_0$ is used here mainly for illustration. Physically, the scaling $f_0\propto r_0^2$ at small $r_0\ll a_0^{-1}$ could arise naturally if the initial 3D distribution of GCs is constant in radius, consistent with the current stellar distribution in Fornax. The peak of the distribution is at $r_0=2a_0^{-1}$, comparable to the current half-light radius.

For the NFW halo, the derivation in \refapp{cdfgcs} shows that before projection effects are taken into account the time-evolved CDF of GC radii at $r\ll r_{\rm cr}\approx1$~kpc takes the form
\be\label{eq:FDtsum} F_{\Delta t}(r)&\approx&A\frac{\tau(r)}{\Delta t},\ee
independent of initial conditions, where $A$ is an $\mathcal{O}(1)$ coefficient.\footnote{For an NFW halo, $A\approx0.4N_{\rm cr}$ where $N_{\rm cr}$ specifies (approximately) the initial number of GCs located inside the critical radius $r_{\rm cr}\approx1$~kpc.} 
This small-$r$ approximation is shown by the dashed red line in the left panel of Fig.~\ref{fig:FDtperp}. 
The dashed green line shows the full unprojected GC CDF, consistent with Eq.~(\ref{eq:FDtsum}) for $r<500$~pc. 

Projection has a significant effect, meaning that typically, a considerable number of GCs observed at projected radius smaller or equal to $r_\perp$ are in fact located at $r>r_\perp$. The solid blue line shows the projected radius CDF: for $r_\perp\lesssim0.5$~kpc, projection roughly doubles the GC count inside a given $r_\perp$. 

The GC timing puzzle is reflected in the left panel of Fig.~\ref{fig:FDtperp}, by the presence of two GCs (GC4 and GC6) inside of $r_\perp<250$~pc or so, where the CDF shown by the solid blue line predicts that no more than one GC should be expected. While Fig.~\ref{fig:FDtperp} shows the CDF resulting from just one example of initial conditions, we could not find initial conditions that would fit the innermost GC4 and GC6, while not at the same time overshooting GC3 and GC2 further out; nor could we find initial conditions which fit GC3, while not at the same time undershooting GC4 and GC6. The reason for this (mild) inconsistency is the growth in $r$ of $F_{\Delta t}$ in Eq.~(\ref{eq:FDtsum}), which is model-independently (that is, irrespective of initial conditions) predicted to possess a strong slope $F_{\Delta t}(r)\propto\tau(r)/\Delta t\propto r^{1.85}$. We stress, however, that the inconsistency is indeed quite mild: the Poisson probability to see two or more GCs where only one GC is expected is about 25\%. This does not seem like severe fine-tuning.

As the right panel of Fig.~\ref{fig:FDtperp} shows, the cored ISO profile can easily provide a time-evolved projected CDF in excellent agreement with the data. While here we performed the calculation for the cored ISO profile, our analysis in this paper makes clear that essentially any model of DM microphysics would lead to similar conclusions as long as it produces a sizable core in Fornax.

Fig.~\ref{fig:FDtperp} suggests that even for a cuspy halo (left panel), the GC timing puzzle does not invoke extreme fine-tuning in the sense that it is not difficult to find apparently reasonable initial conditions that evolve to the observed configuration of GCs. Another point to consider is related to GCs that inspiral down and are tidally disrupted, presumably forming a stellar nucleus \cite{Tremaine1976a}. In the NFW case in Fig.~\ref{fig:FDtperp}, about 50\% of GCs present initially in the halo arrived at the dynamical center within $\Delta t$. 
This may be expected to produce a nucleus for Fornax with a stellar mass in the ballpark of $10^6$~M$_\odot$, which -- as far as we are aware -- is not observed. However, Ref.~\cite{Tremaine1976a} suggested that because of the small number of GCs involved, three-body interactions between accreted GCs could preclude the formation of the nucleus.

More insight could come from numerical simulations, even though both DF and GC formation involve sub-grid physics in most existing simulations. Recently, Ref. \cite{Shao:2020tsl} explored the survival of GCs in cuspy Fornax-like halos using hydrodynamical cosmological simulations \cite{pfeffer2018,kruijssen2019}, to which sub-grid formation of GCs was added and their orbital evolution under DF was tracked in post-processing. 
Ref.~\cite{Shao:2020tsl} calculated the projected radius CDF of GCs at $z=0$, finding that their simulated CDF is consistent with the observed positions of the Fornax GCs. According to Ref.~\cite{Shao:2020tsl}, only around 33\% of GCs are tidally disrupted in the simulations; a somewhat smaller number than the 50\% in our example in Fig.~\ref{fig:FDtperp}. Fornax is found to be special, in that only about 3\% of Fornax-like galaxies in the simulation ended up with five or more surviving GCs today.\footnote{Note that Ref.~\cite{Shao:2020tsl} considered GC6 as a candidate for a tidally disrupted GC, assuming that its projected distance from the center of Fornax is just 30~pc. In comparison, using the Fornax center of Ref.~\cite{wang2019morphology}, we find a projected distance of 254~pc for GC6.}

\section{Summary}\label{sec:summary}
We revisited the calculation of globular cluster orbits under dynamical friction, considering different microscopic models of dark matter and different halo morphologies that they predict. We focused on the Fornax dwarf spheroidal galaxy, which hosts six GCs and which has been noted in previous literature to pose a GC timing problem, that is, the future orbits of some of its GCs are much shorter than their current age.

We presented semi-analytical computations of DF and of GC orbits under DF. For a cuspy DM halo, we showed that the current cumulative distribution function of GC radii takes an approximately power-law form that can be deduced from stellar kinematics and age measurements. Including projection effects, we demonstrated that the GC timing problem does not appear very severe: the existence of the innermost GCs could be accounted for at the cost of moderate fluctuation with a Poisson probability of about 25\%. A comparable hint of a core in Fornax may also be inferred by mass modeling of kinematic data \cite{Read:2018fxs,Hayashi:2020jze}.

A cuspy halo, in conjunction with GC orbits, does place interesting constraints on the initial distribution of GCs in dSphs, as it suggests that an $\mathcal{O}(1)$ of initially present GCs should have arrived and either disrupted or merged at the dynamical center of the galaxy. If GCs merge to form a dense nuclear cluster, as found by some simulations~\cite{CapuzzoDolcetta:2008me,CapuzzoDolcetta:2008jy}, where is the nuclear cluster of Fornax?

Testing these results further would likely require high-resolution numerical simulations including baryonic effects, where GC formation can be modeled from first principles. 
The recent numerical simulations of Ref.~\cite{Shao:2020tsl} made an interesting step in this direction. 
According to Ref.~\cite{Shao:2020tsl} a reasonable distribution of initial conditions for the GCs may naturally lead to the observed present configuration. 
However, both GC formation and DF were treated in Ref.~\cite{Shao:2020tsl} at the sub-grid and post-processing level, and it is not clear (to us) if the resolution of the simulation was high enough to resolve the inner region of Fornax containing the innermost GCs. Moreover, theoretical insight about the role that initial conditions play in shaping the present distribution of GCs is important. We thus believe that our analytical approach remains useful. 

The fact remains that the combination of GC age and orbit measurements could probe the Fornax DM halo and microphysics. 
At the level of the microphysics, in Sec.~\ref{sec:DFfermionic} we calculated DF for three models of DM: fermionic degenerate DM (DDM), where Pauli blocking affects the DF derivation; bosonic ultralight DM (ULDM), where an astronomical de-Broglie scale comes into play; and self-interacting DM (SIDM), where -- in the limit we were mostly interested in -- the microphysics of DF should mostly follow that of CDM, but the halo morphology is different.

For ULDM, DF and specifically the Fornax GC problem were studied in a number of works. Constraints from galaxy dynamics and cosmological Ly-$\alpha$ analyses exclude a soliton core reaching out to the orbits of Fornax GCs, and lead to a similar behavior as in CDM. 

For DDM, we gave a new derivation of DF. We then formulated a robust (in terms of DM model building) version of the Ly-$\alpha$ bound, showing that it excludes an appreciable core, leading again to CDM-like behaviour at the scale of GC orbits. At the same time, stellar kinematics in Fornax could still allow a considerable DDM core. If the Ly-$\alpha$ bound is discounted, for some reason, then DDM could lead to significant suppression of DF and prolongation of the settling time of GC3 and GC4.

For SIDM, stellar kinematics allows a considerable core. If the SIDM cross section is as large as that considered  in Ref.~\cite{Kaplinghat:2015aga}, then the DF settling time for GC3 and GC4 can be significantly longer than in the cuspy halo CDM model.

We also considered the possibility that baryonic feedback deforms a CDM cusp into a core. In that case, the deformation of the halo is expected primarily within the half-light radius \cite{Pontzen2012,Oman2016,Meadows20,Read:2018fxs}. This makes the core spatially smaller than the typical cores that were previously suggested as an explanation of the GC timing puzzle \cite{Goerdt2006,Meadows20}. For our analysis, we adopted the density profile fit in Ref.~\cite{Read:2018fxs}. This intermediate-size baryonic-driven core can also prolong GC orbital decay times within the inner few hundred parsecs compared to the cusp case.

Altogether, we considered DM microphysics (and indirectly, also baryonic feedback) as a possible source for the formation of a core in Fornax, and computed the detailed effects on dynamical friction. In general, both the detailed microphysics and the mere presence of a core (regardless of how it formed) affect the settling of GC orbits. Our analysis suggests that the most relevant factor is the presence of the core itself, rather than the specific microphysics scenario. Further analysis, including other galaxies, and in particular the search for nuclear star clusters in Fornax-like systems, may be able to differentiate between these possibilities.

\acknowledgments
We thank Justin Read for clarifications on the kinematic data and Valerie Domcke, Joshua Eby, Daniel Kaplan and Scott Tremaine for useful discussions. NB is grateful for the support of the Clore scholarship of the Clore Israel Foundation. KB is incumbent of the Dewey David Stone and Harry Levine career development chair at the Weizmann Institute of Science, and was supported by grant 1784/20 from the Israel Science Foundation. HK is supported by the Deutsche Forschungsgemeinschaft under Germany’s Excellence Strategy - EXC 2121 Quantum Universe - 390833306.
%

\begin{appendix}

\section{Dynamical friction in exotic media: derivation from the Boltzmann equation}\label{app:DF}

In this appendix we provide an economical derivation of gravitational DF acting on a nonrelativistic probe object moving in a medium, with different medium microphysics including a classical gas as well as quantum Fermi and Bose gases. We neglect interactions apart from minimal gravity. We start with a quick recap of the derivation of the Fokker-Planck equation, governing the phase-space distribution functions of the probe and medium particles.

We consider the following elastic scattering process of two particle species, 
$$1 (p) + 2 (k) \to 1(p') + 2(k').$$
The phase-space distribution function for the particle species $1$ evolves according to the Boltzmann equation,
\be \frac{df_1}{dt} = C[f_1].\ee
The collision integral $C[f_1]$ contains information about the elastic scattering process, and is written as
\ba
C[f_1] &=& \frac{(2\pi)^4}{2E_p} \int d\Pi_{k} d\Pi_{p'} d\Pi_{k'} \, 
 \delta^{(4)}(p+k - p' - k') |\overline{{\cal M}}|^2 
\nonumber \\
&&
\times \Big[ 
f_1(p') f_2(k') (1 \pm f_1(p) ) ( 1  \pm f_2(k) )\no\\
&& - f_1(p) f_2(k) (1 \pm f_1(p') ) (1 \pm f_2(k') )
\Big],
\ea
where $|\overline{\cal M}|^2$ is a squared matrix element averaged over initial and final spins, and $d\Pi_k = \frac{g}{2E_k} \frac{d^3k}{(2\pi)^3}$ is the Lorentz-invariant phase element with the number of internal degrees of freedom $g$. The sign in $1\pm f_i$ refers to bosons ($+$) or fermions ($-$), respectively. It is convenient to write the above Boltzmann equation in the following form,
\ba
\frac{df_1}{dt} &=& \int \frac{d^3p'}{(2\pi)^3} 
\Big[ 
S(\bfp',\bfp) f_1(p') (1 \pm f_1(p) )\no\\
&-& S(\bfp,\bfp') f_1(p) (1 \pm f_1(p') )
\Big],
\label{be}
\ea
where the function $S$ encodes the response of the medium, and is defined as
\ba\label{eq:Sresp}
S(\bfp,\bfp') &\equiv& \frac{(2\pi)^4}{2E_p 2 E_{p'}}
  \int d\Pi_{k}  d\Pi_{k'} 
 \delta^{(4)}(p+k - p' - k') \times\no\\
&&|\overline{{\cal M}}|^2 
f_2(k) (1 \pm f_2(k') ).
\ea
The function $S(\bfp,\bfp')$ can be interpreted as a differential rate at which a particle of momentum $\bfp$ is converted into a particle with momentum $\bfp'$. 

The Boltzmann equation can be greatly simplified if the momentum exchange
\be q = p' - p\ee
 is smaller than the typical momentum given by the distribution function $f_1$. 
In such cases, the Boltzmann equation is reduced to the nonlinear Fokker-Planck equation,
\be
\frac{df_1}{dt} &=&
- \frac{\partial}{\partial p^i} \left[ f_1 (1\pm f_1) D_i \right]
\\ &+& \frac{1}{2} \frac{\partial}{\partial p^i} \left[ \frac{\partial }{\partial p^j} (D_{ij} f_1) \pm f_1^2 \frac{\partial}{\partial p^j} D_{ij} \right],\no
\ee
where the diffusion coefficients are defined as
\ba
D_i(\bfp) &=&  \int \frac{d^3q}{(2\pi)^3} q^i S(\bfp,\bfp+\bfq),
\label{d1}
\\
D_{ij}(\bfp) &=& \int \frac{d^3q}{(2\pi)^3} q^i q^j S(\bfp,\bfp+\bfq).
\label{d2}
\ea

The gravitational scattering of a probe particle of mass $M$ and a particle in the medium with mass $m$ is described by the spin-averaged matrix element
\bea
| \overline {\cal M}|^2 = \frac{1}{2 s + 1} \frac{(16\pi G)^2 m^4 M^4}{\left[(q^{0})^2-{\bf q}^2\right]^2},
\label{me}
\eea
entering Eq.~(\ref{eq:Sresp}).  
In the nonrelativistic limit, we can neglect $q^0$ and maintain only ${\bf q}$ in Eq.~(\ref{me}).

The problem of calculating the diffusion coefficients for different types of media amounts to evaluating Eqs.~(\ref{d1}) and~(\ref{d2}), where in the response function Eq.~(\ref{eq:Sresp}) we can select the appropriate sign in $1\pm f_2$ corresponding to the medium particle's spin-statistics (or setting $1\pm f_2\to1$ if we wish to compute the classical gas limit).

For the calculation of DF we are particularly interested in $D_{||}$, the first diffusion coefficient corresponding to motion parallel to the  probe object's instantaneous velocity. $D_{||}$ is simply given by Eq.~(\ref{d1}) when we select ${q^i}$ to align with the direction of ${\bf p}$.

\subsection{A classical gas medium}\label{appss:class}
We first re-derive the relaxation of massive classical objects, such as supermassive black holes or GCs, in a background medium consisting of other classical objects such as stars or CDM particles. 
In the nonrelativistic limit, the function $S(\bfp,\bfp')$ is simplified as
\bea
S(\bfp,\bfp') &\simeq &
g_\chi \frac{ (4 \pi G m M)^2}{q^4}
\int \frac{d^3k}{(2\pi)^3} \frac{d^3k'}{(2\pi)^3}
\no\\ && 
\times (2\pi)^4 \delta^{(4)}(p+k-p'-k') f_2(k)
\label{sbh_s}
\eea
where $g_\chi$ is the number of internal degrees of freedom of dark matter. 
Here, $M$ and $m$ are the masses of the particle species $1$ and $2$, respectively. 
In the small momentum exchange limit, the $\delta$-function for the energy conservation can be expanded as
\bea
&&\delta(E_p + E_k - E_{p'} - E_{k'})
\no\\
&&\simeq
\frac{1}{q} \left( 1 + \frac{M}{2\mu_r} \bfq \cdot \frac{\partial}{\partial \bfp} \right) \, \delta\Big[ \hat{q} \cdot \left( \frac{\bfk}{m} - \frac{\bfp}{M}\right) \Big],
\eea
where $\mu_r = m M / (m+M)$ is the reduced mass.
Using these approximate expressions in nonrelativistic and small momentum exchange limit, we obtain the diffusion coefficients as
\bea\label{eq:diffclas}
\!\!\!\! D_i (\bfp)&=&
\int \frac{d^3q}{(2\pi)^3} q^i S(\bfp,\bfp+\bfq)
\\
&=&
4 \pi G^2 m^2 M^2 \left( 1 + \frac{M}{m}\right)
\ln \Lambda
\frac{\partial }{\partial p^i} h(\bfp;\, f_2)\no
\eea
and
\bea
D_{ij} (\bfp) &=& \int \frac{d^3q}{(2\pi)^3} q^i q^j S(\bfp,\bfp+\bfq)
\nonumber\\
&=&
4 \pi G^2 m^2 M^4
\ln \Lambda
\frac{\partial^2}{\partial p^i \partial p^j}
g(\bfp;\, f_2)
\eea
where $\ln \Lambda = \int_{q_{\rm min}}^{q_{\rm max}} dq /q$ is the Coulomb logarithm, and we have used the identities (26)--(27) of Ref.~\cite{Bar-Or:2018pxz} to perform the angular integration at the second step in each equation.
The Rosenbluth potentials $h(\bfp)$ and $g(\bfp)$ are defined as~\cite{Rosenbluth:1957zz}
\bea
h(\bfp;\, f) &=&
g_\chi \int \frac{d^3k}{(2\pi)^3} \frac{ f(k)  }{\big| \frac{\bfk}{m} - \frac{\bfp}{M}\big|},\\
g(\bfp;\, f) &=& g_\chi \int \frac{d^3k}{(2\pi)^3} \Big| \frac{\bfk}{m} - \frac{\bfp}{M} \Big| f(k)  
\eea
This reproduces the well-known diffusion coefficients in a classical system, see Eq. (7.83) in Binney \& Tremaine~\cite{2008gady.book.....B}.
For the Maxwell-Boltzmann distribution $f_2(k) =  (2\pi)^{3/2} n_2 / [g_\chi (m\sigma)^{3}] e^{-v_k^2/2\sigma^2}$, it is straightforward to find
\bea
\frac{\partial h}{\partial p^i} &=& - \frac{v^i}{v} \frac{n_2}{M \sigma^2} \frac{1}{2X^2} \left[ {\rm erf}(X) - \frac{2X}{\sqrt\pi} e^{-X^2} \right]\no\\
&\equiv&  - \frac{v^i}{v} \frac{n_2}{M \sigma^2} G(X),
\\
\frac{\partial g}{\partial p^i\partial p^j} &=& \frac{\sqrt{2}\sigma^2}{M^2}
\Big[ \frac{3}{2} \frac{X^i X^j}{X^3} \Big( G(X) - \frac{1}{3} {\rm erf}(X) \Big) \no\\
&+& \frac{\delta^{ij}}{2} \frac{{\rm erf}(X) - G(X)}{X} \Big],
\eea
where ${\bf v} = \bfp / M$, $v=|{\bf v}|$ and 
\be X&=& \frac{v}{\sqrt{2} \sigma}.\ee

\subsection{Degenerate fermionic dark matter}\label{appss:dfddm}
We now consider the diffusion of astrophysical objects such as GCs in a halo of fermionic dark matter.
In this case, the response function $S$ becomes 
\bea
S(\bfp,\bfp') &\simeq& 
g_\chi \frac{ (4 \pi G m M)^2}{q^4}
\int \frac{d^3k}{(2\pi)^3} \frac{d^3k'}{(2\pi)^3} \,\times\no\\
&& (2\pi)^4\delta^{(4)}(p+k - p' - k') 
f_2(k)  ( 1 - f_2(k') )\no\\&&
\eea
%
Expanding $f_2(k')$ around $k$, we find an additional contribution to the function $S$ due to quantum statistics as
\bea
\Delta S(\bfp,\bfp') &\simeq& - 2\pi g_\chi \frac{ (4 \pi G m M)^2}{q^5}\,\times\\
&&
\int \frac{d^3k}{(2\pi)^3} \left( 1 + \frac{\bfq}{2}\cdot  \frac{\partial}{\partial \bfp} \right)
\delta\Big[ \hat{q} \cdot \bigg( \frac{\bfk}{m} - \frac{\bfp}{M} \bigg) \Big]
f_2^2(k),\no
\label{qc}
\eea
which is the same as Eq.~\eqref{sbh_s} upon substituting $M/\mu \to 1$ and $f_2 \to f_2^2$. 
We find
\bea\label{eq:diffdeg}
D_i(\bfp) &=& 
\frac{4 \pi G^2 m^2 M^3 \ln \Lambda}{\mu_r}  
\frac{\partial}{\partial p^i} \Big[ h(\bfp;\, f_2) - \frac{\mu_r}{M} h(\bfp;\, f_2^2) \Big]
\no\\
\\
D_{ij}(\bfp) &=& 
4 \pi G^2 m^2 M^4
\ln \Lambda
\frac{\partial^2}{\partial p^i \partial p^j}
\Big[ g(\bfp;\, f_2) - g(\bfp;\, f_2^2) \Big]
\no\\
\eea


For the degenerate case, one can perform the $\bfk$ and $\bfk'$ integrations without expanding $f_2(k')$. This computation was already done in the context of neutrino transport in a hot and dense medium~\cite{Reddy:1997yr} and dark matter thermalization in neutron stars~\cite{Bertoni:2013bsa}.
We find:
\bea
S(\bfp,\bfp') &=& 
g_\chi \frac{ (4 \pi G m M)^2}{q^4} 
\frac{m^2 T}{2\pi q} \frac{z}{1- e^{-z}}
\left( 1 + \frac{\xi_-}{z} \right)\no\\&&
\eea
where $z= - q^0/T$, $E_-^2 = m^2 + k_-^2$, $k_-^2 = (m^2/q^2)(q^0 + q^2/2m)^2$, and 
\bea
\xi_- = \ln \left[ \frac{1+ e^{(E_- -\mu)/T}}{1+ e^{(E_- -\mu)/T }e^z } \right].
\eea
Integrating this response function with respect to ${\bf q}$, one obtains the diffusion coefficients for a degenerate medium.  

\subsection{Ultralight dark matter}\label{appss:dfuldm}
It was discussed in~\cite{Hui:2016ltb} that the dynamical relaxation of stars in a ULDM halo proceeds as stars scatter off ULDM quasi-particles whose size is of the order of the de Broglie wavelength, $\lambda_{\rm dB} \sim 2\pi/mv$. This observation was confirmed by Bar-Or et al.~\cite{Bar-Or:2018pxz}, where the dynamical relaxation time scale as well as diffusion coefficients were computed in a more rigorous way by using Fokker-Planck equation and stochastic gravitational potential. 

The Boltzmann equation approach can also reproduce the dynamical relaxation time scale and diffusion coefficients.
The gravitational scattering between ultralight dark matter and a star can be described by the same matrix element, Eq.~\eqref{me}, where $M$ and $m$ are the mass of the star and ultralight dark matter, respectively. We treat the star as a pointlike particle, and this can be justified since the maximum  momentum exchange $q \sim m v$ is much smaller than $1/r$ with a typical star radius $r$. 
The function $S(\bfp,\bfp')$ is
\bea
S(\bfp,\bfp') &\simeq &
g_\chi \frac{ (4 \pi G m M)^2}{q^4}
\int \frac{d^3k}{(2\pi)^3} \frac{d^3k'}{(2\pi)^3}  \,\times\no\\&&
(2\pi)^4 \delta^{(4)}(p+k - p' - k') 
f_2(k)  ( 1 + f_2(k') ).\no\\&&
\eea
The quantum correction is the same as Eq.~\eqref{qc} with an opposite sign. 
Therefore, the diffusion coefficients are
\bea
D_i(\bfp) &=& 
\frac{4 \pi G^2 m^2 M^3 \ln \Lambda}{\mu_r}
%
\frac{\partial}{\partial p^i} 
\Big[ h(\bfp;\, f_2) + \frac{\mu_r}{M} h(\bfp;\, f_2^2) \Big]
\no\\
\\
D_{ij}(\bfp) &=& 
4 \pi G^2 m^2 M^4
\ln \Lambda
\frac{\partial^2}{\partial p^i \partial p^j}
\Big[ g(\bfp;\, f_2) + g(\bfp;\, f_2^2) \Big]
\no\\
\eea
This reproduces the results of Ref.~\cite{Bar-Or:2018pxz}.

\section{Maximum entropy DDM halos}\label{app:profile}

In the derivation of the quasi-degenerate density profile, we adopt the assumption that a galactic structure may be described as a statistical ensemble close to equilibrium, in the sense of a maximal Boltzmann-Gibbs entropy. A similar approach to ours can be found in a number of earlier works \cite{lyndenbell67,lyndenbell68,Chavanis:2002rj,Chavanis:2002yv,Chavanis_2004,Chavanis:2014xoa,Domcke2015}. 

The phase-space distribution function $f({\bf r},{\bf p})$ and differential particle number density $dN$ are related via
\be\label{eq:dNd3rd3p}
(2\pi)^3\frac{dN}{d^3xd^3p} = g f(\mathbf{r},\mathbf{p}) \; .
\ee
The entropy of the gas is then given by the functional
\be
S = -g\int \frac{d^3pd^3r}{(2\pi)^3} \left[f\ln f+(1-f)\ln(1-f)\right] \; .
\ee
Supplemented with Lagrange multipliers for the total energy and the total number of particles, the variation problem can be carried out along the lines of Ref.~\cite{lyndenbell68}. The maximum entropy result is
\be\label{eq:fasz}
f(\mathbf{r},\mathbf{p}) = \frac{1}{1+\exp[z(\mathbf{r},\mathbf{p})]} \; ,
\ee
where
\be
z(\mathbf{r},\mathbf{p}) = \frac{\beta \mathbf{p}^2}{2m} + \beta m\Phi(\mathbf{r}) + \alpha \; ,
\ee
with $ \beta $ and $ \alpha $ being the energy and particle number Lagrange multipliers. The gravitational potential is given by
\be
\Phi(\mathbf{r}) &=& -Gm g\int \frac{d^3p^{\prime}d^3r^{\prime}}{(2\pi)^3}\frac{f(\mathbf{r}^{\prime},\mathbf{p}^{\prime})}{|\mathbf{r}-\mathbf{r}^{\prime}|} \; .
\ee
By construction, the gravitational potential solves the Poisson equation, $ \nabla^2\Phi = 4\pi G \rho $, where the density is
\be
\rho(\mathbf{r}) &=& mg\int \frac{d^3p}{(2\pi)^3} f(\mathbf{r},\mathbf{p}) \; .
\ee
%
We can make progress by evaluating the density,
\be
\rho(\mathbf{r}) &=& -mg\left(\frac{m}{2\pi \beta}\right)^{3/2}{\rm PolyLog}\left[\frac{3}{2},-e^{\varphi} \right] \; ,
\ee
where $ \varphi \equiv -\beta m \Phi(\mathbf{r})-\alpha \equiv \beta \mu=\mu/T$, defining also the ``chemical potential'' $ \mu(\mathbf{r}) = -m\Phi(\mathbf{r})-\widetilde{\alpha} $, with $ \widetilde{\alpha} \equiv \alpha/\beta $. 

Using the Poisson equation and the definition of $ \varphi $, we have
\be\label{eq:phipoisson}
\nabla^2\Phi = -\frac{1}{\beta m} \nabla^2\varphi = 4\pi G \rho \; .
\ee

In the degenerate limit, $ \mu/T =\varphi \gg 1 $, the PolyLog function asymptotes to
\be
-{\rm PolyLog}\left[\frac{3}{2},-e^{\varphi} \right] \to \frac{4\varphi^{3/2}}{3\sqrt{\pi}} =\frac{4( \mu/T)^{3/2}}{3\sqrt{\pi}} \; . 
\ee
It is therefore useful to rewrite Eq.~\eqref{eq:phipoisson} as
\be
&\nabla^2& \mu(\mathbf{r}) \\&=& \frac{4\sqrt{2}}{3\pi}g G m^{7/2}\mu_0^{1/2} \frac{{\rm PolyLog}\left[\frac{3}{2},-e^{\beta\mu_0 f}\right]}{\frac{4}{3\sqrt{\pi}}(\beta \mu_0)^{3/2}}  \; ,\no
\ee
where we defined $\mu(0)\equiv\mu_0$ and $ \mu(\mathbf{r})=\mu_0 h(\mathbf{r}) $. Let us also define a dimensionless radius $x$ via $r= r_0x $, with $r_0$ given by
\be
r_0&=& \sqrt{\frac{3\pi}{4\sqrt{2}g Gm^{7/2}\mu_0^{1/2}}}\; .
\ee
Then, one finds the equation
\be\label{eq:fdiffeq}
\partial_x (x^2\partial_x h) = x^2 \frac{{\rm PolyLog}\left[\frac{3}{2},-e^{(\mu_0/T) h}\right]}{\frac{4}{3\sqrt{\pi}}(\mu_0/T)^{3/2}} \; .
\ee
Given the solution for $h$, the density is simply
\be\label{eq:rhowithf}
\rho &=& \frac{\sqrt{2}}{3\pi^2}g m^{\frac{5}{2}}\mu_0^{\frac{3}{2}}\frac{-{\rm PolyLog}\left[\frac{3}{2},-e^{\frac{\mu_0}{T} h}\right]}{\frac{4}{3\sqrt{\pi}}(\mu_0/T)^{3/2}} \; .
\ee

\begin{figure}[htbp!]
	\centering
	\includegraphics[width=0.5\textwidth]{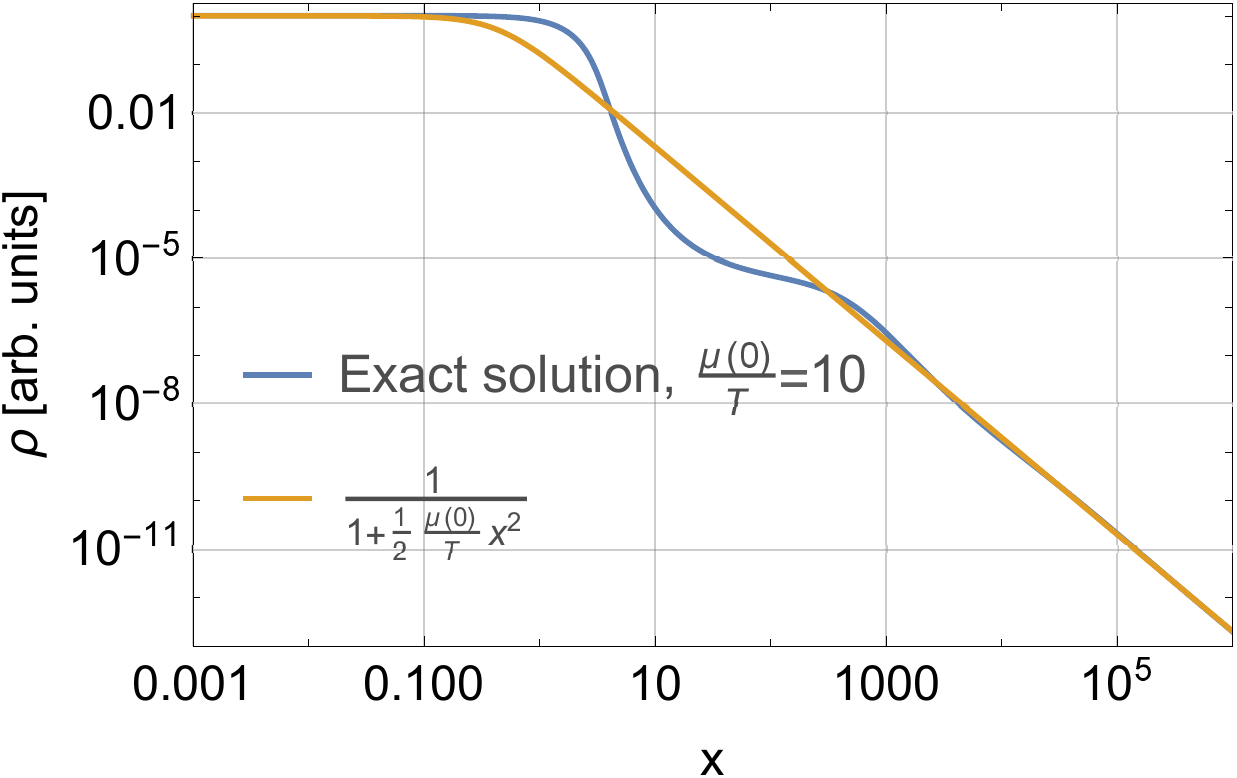}	
	\caption{The density profile found by solving Eq.~\eqref{eq:fdiffeq} and inserting into Eq.~\eqref{eq:rhowithf} (blue line), compared with an analytical ansatz (orange) that demonstrates the asymptotic behavior of the solution. Evidently, for $ x\lesssim 1 $ the density profile is constant, whereas the density asymptotes to $ 1/x^2 $ for large $ x $.}\label{fig:densityMaxEnt}
\end{figure}

In the limit $ \mu_0/T\gg 1 $, the right-hand side of Eq.~(\ref{eq:fdiffeq}) becomes $ -x^2 h^{3/2} $, which is a Lane-Emden equation with a scale $ r_0 $. The central density becomes 
$\rho(0)\equiv\rho_0\approx (\sqrt{2}/3\pi^2)gm^{5/2}\mu_0^{3/2} $, 
i.e. $ \mu_0=(3\pi^2/\sqrt{2})^{2/3}(\rho_0/g)^{2/3}/m^{5/3} $, 
which implies\footnote{Our natural unit notation, which includes $\hbar\to1$, may mask the fact that the characteristic radius Eq.~(\ref{eq:ddmr0}) is determined by quantum degeneracy pressure. This is easy to unmask by restoring $(2\pi)^3\to(2\pi\hbar)^3$ on the left-hand side of Eq.~(\ref{eq:dNd3rd3p}). Tracking $\hbar$ through the computation gives a factor of $\hbar$ on the right-hand side of Eq.~(\ref{eq:ddmr0}).} 
\be\label{eq:ddmr0} r_0 &\approx& \frac{1}{2}\left(\frac{9\pi}{2G^3\,\rho_0\,g^{2}m^{8}}\right)^{\frac{1}{6}}.
\ee
The solution for $\rho$ is constant near the origin and falls as $\rho\propto 1/r^2$ at $r\gg r_0$, with a ``wriggle" feature near $r\sim r_0$.   
An example with $\mu_0/T=10$ is shown in Fig.~\ref{fig:densityMaxEnt}.

\begin{figure}[htbp!]
	\centering
	\includegraphics[width=0.45\textwidth]{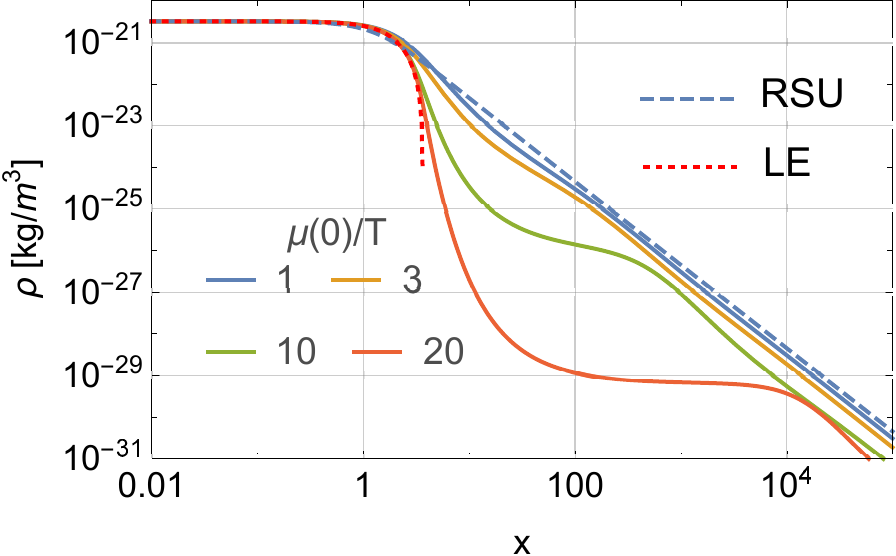}	
	\caption{The density profile found by solving Eq.~\eqref{eq:fdiffeq} and inserting into Eq.~\eqref{eq:rhowithf} for different values of $ \mu_0/T  $, keeping $ \rho_0 $ constant. The dashed blue line (RSU) is based on the treatment of Ref.~\cite{Randall2017}. The dotted red line (LE) is based on solving the Lane-Emden equation, which is the $ \mu_0/T\to \infty $ limit of the equations, as described in the text.}\label{fig:densityDiffmuT}
\end{figure}

It is interesting to compare this class of solutions to the solutions obtained from the prescription of Ref.~\cite{Randall2017}. Rescaling the equations of Ref.~\cite{Randall2017} by $ r_{0} $ from Eq.~(\ref{eq:ddmr0}), we plot the solution we find (at constant $ \rho_0 $) in Fig.~\ref{fig:densityDiffmuT} (named RSU), along with different solutions corresponding to different $ \mu_0/T $. Evidently, the profile used in Ref.~\cite{Randall2017} bears a strong resemblance to the $ \mu_0/T\sim 1 $ case. Also, Fig.~\ref{fig:densityDiffmuT} shows that in the limit $ \mu_0/T\to \infty $, the solution is a core with finite radius. We show it both by solving the density profile for $ \mu_0/T\gg 1 $ and by solving the Lame-Emden (LE) approximation that appears above.

\begin{figure}[htbp!]
	\centering
	\includegraphics[width=0.44\textwidth]{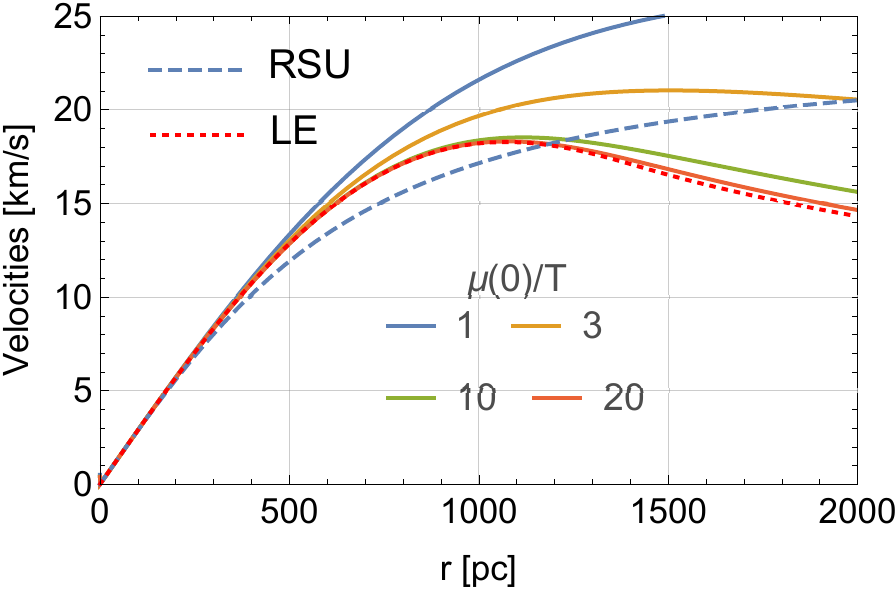}	
	\caption{The circular velocities $ \sqrt{GM(r)/r} $ for the density profiles that appear in Fig.~\ref{fig:densityDiffmuT}.}\label{fig:velocityDiffmuT}
\end{figure}

In Fig.~\ref{fig:velocityDiffmuT} we plot the circular velocities induced by the density profiles in Fig.~\ref{fig:densityDiffmuT}.


\section{Jeans modeling}\label{app:jeans}
Following Ref.~\cite{BinneyTremaine2}, the equation for the second velocity moments of a static spherical system of particles under the influence of a gravitational potential with enclosed mass $ M(r) $  is
\be\label{eq:jeans}
\frac{1}{\nu}\frac{d}{dr}(\nu \bar{v^2_r})+2\frac{\beta \bar{v^2_r}}{r} = -\frac{GM}{r^2} \; ,
\ee
where $ \nu(r) $ is the particles' density, $ \bar{v^2_r}(r) $ is the radial second velocity moment, $ \bar{v^2_\theta} $ is the angular second velocity moment and $ \beta \equiv 1-\bar{v^2_\theta}/\bar{v^2_r} $ is the velocity anisotropy. For constant $ \beta $, Eq.~\eqref{eq:jeans} is solved by
\be\label{eq:jeanslight}
\nu \bar{v^2_r}(r) &=& \frac{G}{r^{2\beta}}\int\limits_r^{\infty} r^{\prime 2\beta-2}\nu(r^{\prime})M(r^{\prime})dr^{\prime} \; .
\ee
This can be related to the line-of-sight (LOS) velocity,
\be
\sigma^2_{\rm LOS}(r) &=& \frac{2}{I(r)}\int\limits_r^{\infty}\left(1-\beta\frac{r^2}{r^{\prime 2}}\right) \frac{\nu \bar{v^2_r}(r^{\prime}) r^{\prime}}{\sqrt{r^{\prime 2}-r^2}}dr^{\prime} \; .
\ee
In modeling stellar kinematics in Fornax, we assume a Plummer profile with density and surface density given by, respectively,
\be
\nu(r) &=& \frac{1}{(1+r^2/r_p^2)^{5/2}}\frac{3L}{4\pi r_p^3} \\
I(r) &=&  \frac{1}{(1+r^2/r_p^2)^{2}}\frac{L}{\pi r_p^2} \; .
\ee
We use the radius parameter $ r_p=710 ~$pc \cite{Read:2018fxs}, consistent with the stellar sample on which the kinematics data is based. (This radius parameter is about 20\% smaller than the $ r_p \approx 851 $~pc reported in a new morphological study \cite{wang2019morphology}. The difference is not crucial for our analysis. Moreover, we prefer to consider the photometry and spectroscopy of the same data set.) 

\section{CDM velocity dispersion in a cored profile}\label{app:cdm}


In this appendix we discuss some features of cored CDM halos, notably DF, partially following Ref.~\cite{Petts2015}.

The velocity dispersion of dark matter is important to the discussion, therefore let us write the Jeans equation for the second radial velocity moment with a constant velocity anisotropy $ \beta $, which has the solution \cite{BinneyTremaine2}
\be\label{eq:veljeans}
\bar{v^2_r}(r) = \frac{G}{r^{2\beta}\rho(r)}\int\limits_r^{\infty} r^{\prime 2\beta}\frac{\rho(r^{\prime})M(r^{\prime})}{r^{\prime 2}} dr^{\prime} \; ,
\ee
which is the same as \refeq{jeanslight} but in slightly different notation.

Consider a finite-core toy model, where the density is $ \rho(0) $ for $ r<r_{\rm c} $ and $ 0 $ for $ r> r_{\rm c}$. Then, the solution of Eq.~\eqref{eq:veljeans} is \cite{Petts2015}
\be
\bar{v^2_r}(r) = \frac{2\pi G\rho(0)}{3(\beta+1)}\frac{1}{r^{2\beta}}\left(r_{\rm c}^{2\beta+2}-r^{2\beta+2}\right) \; . 
\ee
For isotropic velocity dispersion $ \beta=0 $, this reduces to
\be
 \bar{v^2_r}(r;\beta=0) = \frac{2\pi G\rho(0)}{3}\left(r_{\rm c}^{2}-r^{2}\right) \; . 
\ee
Thus, for $ r\ll r_{\rm c} $,
\be\label{eq:sigrFC}
\sigma_r & \equiv & \sqrt{\bar{v^2_r}(r\ll r_c;\beta=0)} \approx\sqrt{\frac{2\pi G\rho(0)}{3}}r_{\rm c} \\ & \approx &  30\left(\frac{\rho(0)}{ 10^{8}~\frac{M_{\odot}}{{\rm kpc}^3}}\right)^{\frac{1}{2}}\frac{r_{\rm c}}{1~{\rm kpc}}\frac{{\rm km}}{{\rm s}} \; . \no 
\ee
We can also note the ratio,
\be\label{eq:voversigcdmcore}
X\equiv \frac{V_{\rm circ}}{\sqrt{2}\sigma_r}\approx \sqrt{\frac{GM(r)}{2r}}\Big/\sqrt{\frac{2\pi G\rho(0) r_c^2}{3}} = \frac{r}{r_c} \; ,
\ee
indicating that the low-velocity approximation of the Chandrasekhar deceleration may apply inside a core, see Eq.~\eqref{eq:CMax}. This implies a ``phase-space suppression'' to DF, as discussed in the main text (Sec.~\ref{s:cdmcorecusp}).

\section{Orbits under dynamical friction}\label{app:orbits}
In this appendix we review the solution of an orbit under the influence of DF. 
We write the equations of motion (EoM) in circular coordinates,
\be
\ddot{\mathbf{r}} & = & (\ddot{r}-r\dot{\varphi}^2)\hat{r} + (2\dot{r}\dot{\varphi}+r\ddot{\varphi})\hat{\varphi} \\ &=& -\frac{GM(r)}{r^2}\hat{r} - \Big|\frac{d\dot{\mathbf{r}}}{dt}\Big|_{\rm DF}\frac{\dot{\mathbf{r}}}{|\dot{\mathbf{r}}|} \; .
\ee
We express the deceleration $ |d\dot{\mathbf{r}}/dt|_{\rm DF} $ as $ |\dot{\mathbf{r}}|/\tau $, where $ \tau $ appears in Eq.~\eqref{eq:genDecaytime2}.

Defining $ r=R_0 x $, $ t=T_0\bar{t}  $, $ T_0^2 = R_0^3/GM(R_0) $, we find
\be
x^{\prime\prime}-x\varphi^{\prime 2} &=& -\frac{1}{x^2}\frac{M(R_0x)}{M(R_0)} - \frac{x^{\prime}}{\tau/T_0} \\ 2x^{\prime}\varphi^{\prime} +x\varphi^{\prime\prime} & = &  - \frac{x \varphi^{\prime}}{\tau/T_0}\label{eq:angulareom}
\ee
where $ \prime $ is differentiation with respect to $ \bar{t} $. Note, $ \tau $ can depend on $ r$ and $ |\dot{\mathbf{r}}| $. For a circular orbit, for example, the initial conditions can be set as $ x(0)=1 $, $ x^{\prime}(0)=0 $, $ \varphi(0)=0 $ and $ \varphi'(0)=1/x(0)= 1 $, 
which has a revolution time of $ \Delta \widetilde{t}=2\pi $. 

Solving the orbit of a decelerating test object generally requires numerical integration. We can understand some features of the solution analytically, however. Defining $ v_\varphi \equiv r\dot{\varphi} $, the $ \hat{\varphi} $ part of the EoM has the solution
\be\label{eq:eomang}
r v_\varphi  = (r v_\varphi)_0\exp\left(-\int\limits_0^t\frac{dt^{\prime}}{\tau}\right) \; .
\ee
This solution expresses the decay of angular momentum of the test object. Using the circular velocity $ v_{\rm circ}^2=GM(r)/r  $, we can express the $ \hat{r} $ part of the EoM as
\be\label{eq:reomvels}
v_\varphi^2-v_{\rm circ}^2 = r\left(\ddot{r}+\frac{\dot{r}}{\tau}\right) \; .
\ee

We can gain more analytical intuition by considering nearly circular orbits, assuming that the inspiral rate is much smaller than the circular velocity, $r/\tau\ll v_{\rm circ}$.
 Assuming that $ \dot{r}\sim r/\tau $, $ \ddot{r}\sim r/\tau^2 $ and $ r/\tau \ll v_{\rm circ} $,
 \refeq{reomvels} implies $ v_\varphi\approx v_{\rm circ} $. 
 We can use this to write
 \be
 -\frac{r v_\varphi}{\tau} &=& \dot{r} v_\varphi + r\dot{v}_\varphi \approx \dot{r} v_{\rm circ}+r\dot{v}_{\rm circ} \\ & = & \frac{1}{2}v_{\rm circ} \dot{r} \left(1+\frac{d\ln M}{d\ln r}\right) \; .
 \ee
Rearranging, we find
\be\label{eq:rdotrtau}
\frac{\dot{r}}{r} \approx - \frac{2}{\left(1+\frac{d\ln M}{d\ln r}\right)\tau} \; .
\ee
Using this, we can estimate the time it takes a test object to fall from $ r_0 $ down to $ r<r_0 $:
\be\label{eq:trr0}
t(r; r_0) = \int\limits_r^{r_0}\frac{dr}{2r}\left(1+\frac{d\ln M}{d\ln r}\right) \tau(r,v_{\rm circ}(r)) \; .
\ee
Given the mass profile of the halo, $ M(r) $, and a DF model encapsulated by $ \tau $, \refeq{trr0} is a simple and quick estimate of the inspiral time of a test object.

For eccentric orbits, the approximation above is less justified. Defining eccentricity as $ e\equiv (r_{\rm apo}-r_{\rm peri})/(r_{\rm apo}+r_{\rm peri}) $ with apocenter radius $ r_{\rm apo} $ and pericenter radius $ r_{\rm peri} $, we numerically tested \refeq{trr0} for $ e>0 $. In these calculations we defined $ r_0 $ and $ r $ via $ (r_{\rm apo}+r_{\rm peri})/2 $, where $ r_{\rm apo} $ and $ r_{\rm peri} $ are obtained per cycle of the orbital phase. With these definitions, in numerical experiments representative of Fornax GCs we find that \refeq{trr0} holds to better than $ 30\% $ accuracy for $ e\lesssim 0.5 $. 


\section{The radial and projected CDF of GCs}\label{app:cdfgcs}
%

Consider a population of identical GCs (all with the same mass), that start off their life at some initial time $t=0$ on approximately circular orbits with a radial probability distribution function (PDF) $ f_0(r_0) $ w.r.t. an initial radial coordinate $r_0$. The CDF of initial GC positions is $F_0(r_0)=\int_0^{r_0}dyf_0(y)$. We are interested in computing the PDF and CDF of GC radial positions today, at $ t=\Delta t$; call these $f_{\Delta t}(r)$ and $F_{\Delta t}(r)$.

DF causes GC orbits to inspiral inwards, and by integrating along the orbit we can compute the function $r=r(r_0;\Delta t)$ and invert it to obtain $r_0=r_0(r;\Delta t)$.\footnote{The monotonous decrease of $r$ with time, that allowed this inversion, is lost for non-circular orbits. We could accommodate elliptical orbits approximately, by letting $r$ represent the average between the peri- and apo-center per cycle.} 
Neglecting tidal disruption, we have 
%
\be\label{eq:cdf}
F_{\Delta t}(r) 
&=&F_0(r_0(r;\Delta t))\; .
\ee
Now we can use explicit results for $r_0(r;\Delta t)$ to connect $F_{\Delta t}(r)$ with $F_0(r_0)$ in different halo models.  
To this end we can use \refeq{trr0},
\be\label{eq:deltat}
\Delta t = \int\limits_r^{r_0}\frac{dr^{\prime }}{2r^{\prime}}\left(1+\alpha(r^{\prime})\right) \tau(r^{\prime}) \; ,
\ee
where $ \alpha(r) \equiv d\ln M/d\ln r $. Let us consider the general features of $ F_{\Delta t} $ for different halo shapes.

\subsection{CDF of GCs in a cuspy halo}
%
We have seen in the main analysis that a cuspy halo (i.e. the inner region of an NFW halo, where $\alpha\approx2$) exhibits an approximately power-law form for the DF time $\tau$. 
For an approximately constant $ \alpha $ and power law $ \tau = \bar{\tau}(r/\bar{r})^{\beta} $, it is useful to define the {\it critical radius} $r_{\rm cr}$ via
\be\tau(r_{\rm cr})&=&\frac{2\beta}{1+\alpha}\Delta t.\ee
The physical meaning of $r_{\rm cr}$ is that GCs that start their life at $r_0\leq r_{\rm cr}$ arrive at the origin within $t\leq\Delta t$. Using our power-law form for $\tau$, we have
\be \label{eq:rcr}r_{\rm cr}&=&\bar{r}\left(\frac{2\beta}{1+\alpha}\frac{\Delta t}{\bar{\tau}}\right)^{1/\beta}.\ee
In terms of $r_{\rm cr}$, the solution of \refeq{deltat} evaluates to
\be
r_0(r;\Delta t) &=& r_{\rm cr}\left(1+\left(\frac{r}{r_{\rm cr}}\right)^\beta\right)^{1/\beta} \; .
\ee

For GCs that satisfy $r\ll r_{\rm cr}$ today, we can expand their starting point:
\be\label{eq:r0exp} r_0(r;\Delta t)&=&r_{\rm cr}+\frac{r_{\rm cr}}{\beta}\left(\frac{r}{r_{\rm cr}}\right)^\beta+...\\
&=&r_{\rm cr}\left(1+\frac{1+\alpha}{2\beta^2}\frac{\tau(r)}{\Delta t}+...\right)\;.\no\ee
In other words, for cuspy CDM halos, GCs that are currently seen at $r\ll r_{\rm cr}$ must have originated near $r_0\approx r_{\rm cr}$. This means that for GCs with $r\ll r_{\rm cr}$ today, the radial distribution today is not very sensitive to the (difficult to predict) initial distribution. We can make this point manifest by expanding Eq.~(\ref{eq:cdf}), using Eqs.~(\ref{eq:rcr}) and~(\ref{eq:r0exp}):
\be\label{eq:FDt} F_{\Delta t}(r)&\approx&F_0(r_{\rm cr})+\frac{(1+\alpha)}{2\beta^2}f_0(r_{\rm cr})r_{\rm cr}\frac{\tau(r)}{\Delta t}+...\;,\ee
where the $...$ refer to higher powers of the small ratio $\tau(r)/\Delta t$. Above, the $r$-independent constant $F_0(r_{\rm cr})$ counts GCs that have already settled to the center of the halo. These  GCs at $r\approx0$ were likely tidally disrupted, suggesting that in actually counting GCs in the system, we should eliminate the term $F_0(r_{\rm cr})$ on the RHS of Eq.~(\ref{eq:FDt}). We thus have $F_{\Delta t}(r)\approx A\left(\tau(r)/\Delta t\right)$, where $A=\frac{(1+\alpha)}{2\beta^2}f_0(r_{\rm cr})r_{\rm cr}$ is an order-unity constant ($r$-independent) coefficient.\footnote{For the inner region of an NFW profile, we have seen that $\alpha\approx2$ and $\beta\approx2$, so $A\approx 0.4N_{\rm cr}$, where $N_{\rm cr}=f_0(r_{\rm cr})r_{\rm cr}$ counts the number of GCs that were located in a region of order $r_{\rm cr}$ around $r_{\rm cr}$. 
Predicting the actual value of $N_{\rm cr}$ would require understanding of the initial cosmological formation of GCs, which is still not under full theoretical control.} 

We can summarize this section with two important conclusions. First, for a cuspy halo, all GCs that are born at $r<r_{\rm cr}$ have arrived at $r\approx0$ by today and are plausibly tidally disrupted. This means that observations today are not sensitive to initial conditions, characterized by different $f_0(r_0)$, that differ from each other only at $r<r_{\rm cr}$; unless stellar age and metallicity measurements can identify  the remnants and approximately count tidally disrupted GCs, on time scales of Gyrs after the disruption. Second, the radial CDF of GCs at small radii $r\ll r_{\rm cr}$ should follow $F_{\Delta t}(r)\approx A\left(\tau(r)/\Delta t\right)$, with order-unity $A$, irrespective of initial conditions. 
\begin{figure*}[htbp!]
	\centering
	\includegraphics[width=0.45\textwidth]{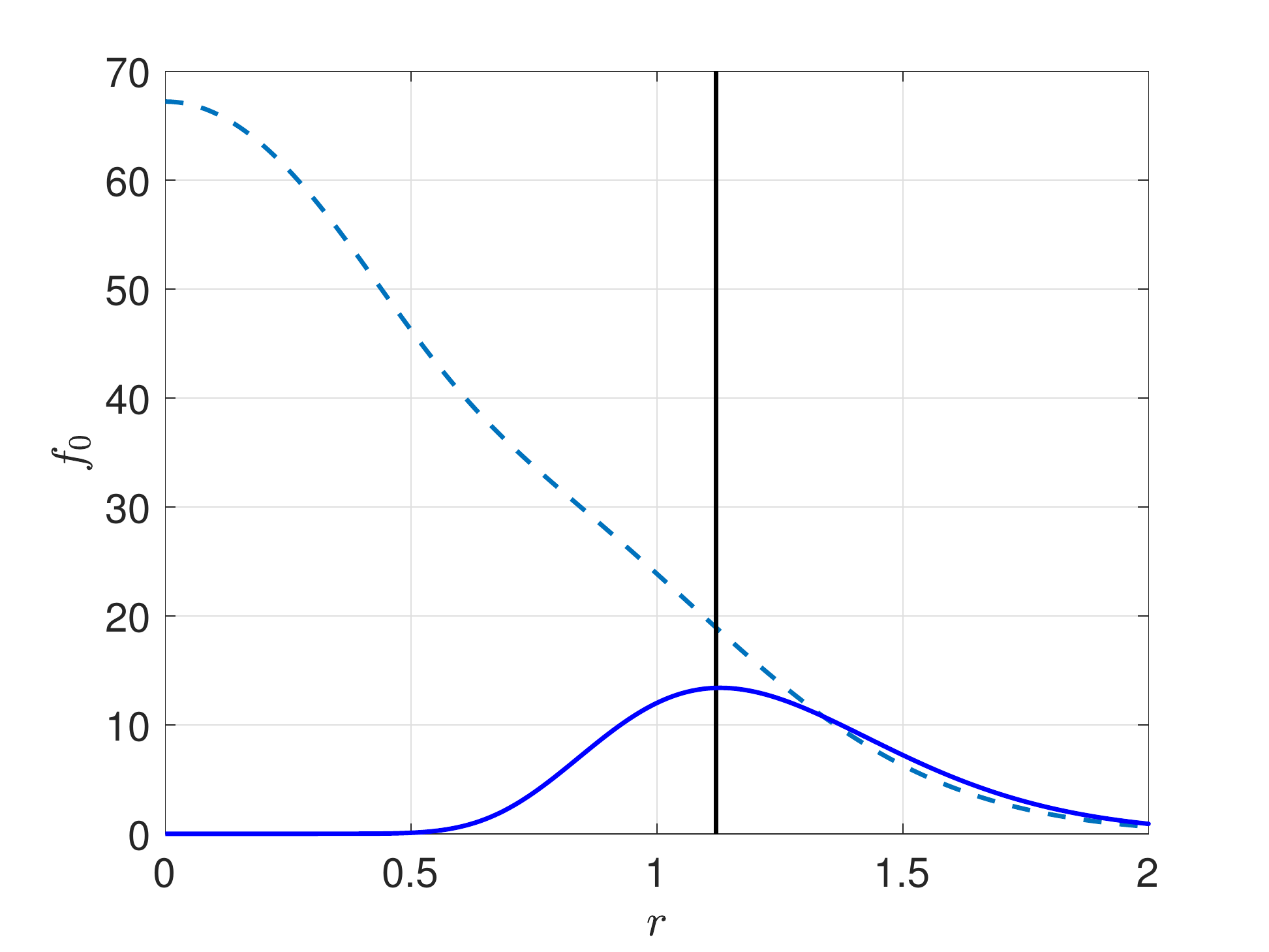}
	\includegraphics[width=0.45\textwidth]{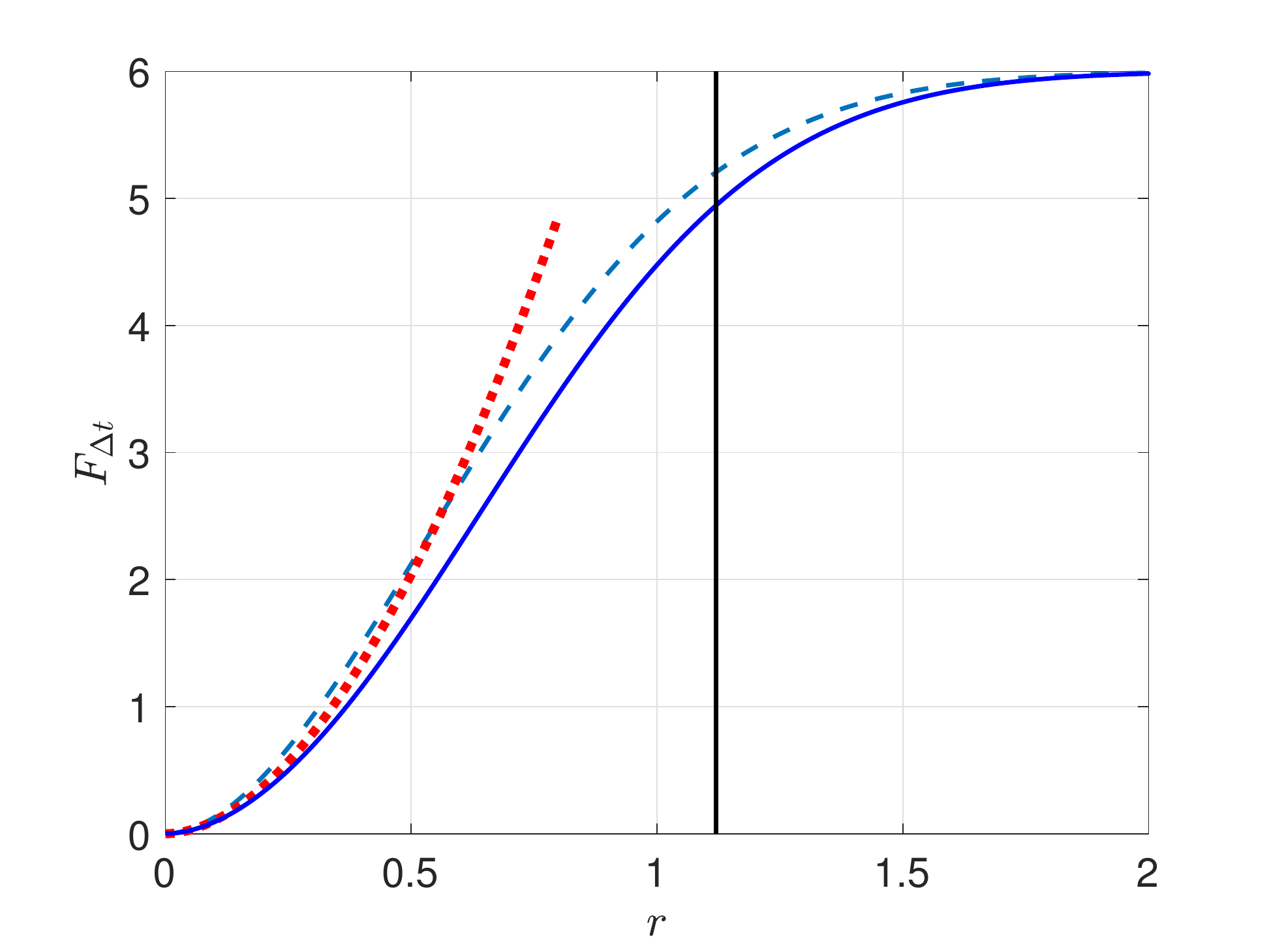}
	\caption{Initial radial PDF $f_0$ ({\bf left}) and resulting current radial CDF $F_{\Delta t}$ ({\bf right}) for a cuspy halo with $\tau(r)\propto r^{1.85}$ and $r_{\rm cr}=1.12$ (units on the x-axis are arbitrary). Two different examples for $f_0$ are shown, leading to nearly identical $F_{\Delta t}$. The two versions of $f_0$ are normalized to yield $F_{\Delta t}(\infty)=6$. In both panels, $r_{\rm cr}$ is marked with a vertical black line. On the left, the $r\ll r_{\rm cr}$ approximation $F_{\Delta t}\approx A\left(\tau(r)/\Delta t\right)$ is shown by the red dotted curve.}\label{fig:PDFCDF}
\end{figure*}

Fig.~\ref{fig:PDFCDF} illustrates both of these two points, by showing two examples of $f_0$ and the resulting $F_{\Delta t}$. The DF time $\tau(r)$ and the critical radius $r_{\rm cr}$ are measurable given a model of the DM halo, fitted to stellar kinematics, and given GC age measurements that define $\Delta t$. This makes the predicted shape of $F_{\Delta t}$ measurable, in principle. In practice, however,  projection effects (explained below) complicate the interpretation. In addition, the collection of GCs in Fornax seems too sparse to draw robust conclusions.

\subsection{CDF of GCs in a cored halo}
Inside a core we expect $ \alpha\approx 3 $ and an approximately constant $ \tau$. \refeq{deltat} is evaluated to
\be
\Delta t = \frac{1+\alpha}{2}\tau \ln \frac{r_0}{r} 
\ee
and the radial CDF today is 
\be
F_{\Delta t}(r) &\approx& F_0(r e^{\frac{2\Delta t}{(1+\alpha)\tau}})\; \approx F_0(r e^{\frac{\Delta t}{2\tau}}) \; .
\ee
The distribution of GCs inside a core reflects a stretched version of the initial conditions. Because of this sensitivity to initial conditions, the degree of possible fine-tuning in the current positions of GCs may be difficult to assess.


\subsection{Accounting for distribution of GC masses}
The masses of GCs in Fornax vary over about an order of magnitude around $10^{5}~M_{\odot} $, and the instantaneous DF time satisfies $\tau\propto1/m_*$ up to logarithmic corrections that we neglect here. 
It is therefore necessary to revise the prediction of the radial CDF of GCs to account for different GC masses. 

In the case of a cuspy profile, where $\tau\propto r^\beta$, the critical radius scales as $r_{\rm cr}\propto m_*^{1/\beta}$.  
For example, using the NFW fit of Fornax (which gives $\beta\approx1.85$), relevant GC masses, and $\Delta t=12$~Gyr we have:
\be
r_{\rm cr} & \approx & 0.7 \left(\frac{\M}{M_{\rm GC4}}\right)^{0.54}~{\rm kpc} \\ & \approx & 1.6 \left(\frac{\M}{M_{\rm GC3}}\right)^{0.54}~{\rm kpc} \; .
\ee

Suppose we have a set of GC masses $m_{*i}$ with initial radial distribution functions $f_{0,i}(r_0)$. Summing over all GC masses we find that as long as $r\ll r_{{\rm cr},i}$, Eq.~(\ref{eq:FDt}) predicts that the total radial CDF today is (again omitting GCs that have already settled to the center of the halo)
%
\be\label{eq:FDtm} \sum_iF_{\Delta t,i}(r)
%
&\approx&F_{\Delta t,1}(r)\sum_i\frac{f_{0,i}(r_{{\rm cr},i})}{f_{0,1}(r_{{\rm cr},1})}\left(\frac{m_{*i}}{m_{*1}}\right)^{\frac{1}{\beta}-1}\;,\no\\&&
\ee
where $F_{\Delta t,1}(r)$ is the radial CDF of GCs of mass $m_{*1}$.
We see that Eq.~(\ref{eq:FDtm}) simply reproduces Eq.~(\ref{eq:FDt}) up to a modified overall multiplicative constant.

\subsection{Projected radius distribution}
\begin{figure}[htbp!]
	\centering
	\includegraphics[width=0.45\textwidth]{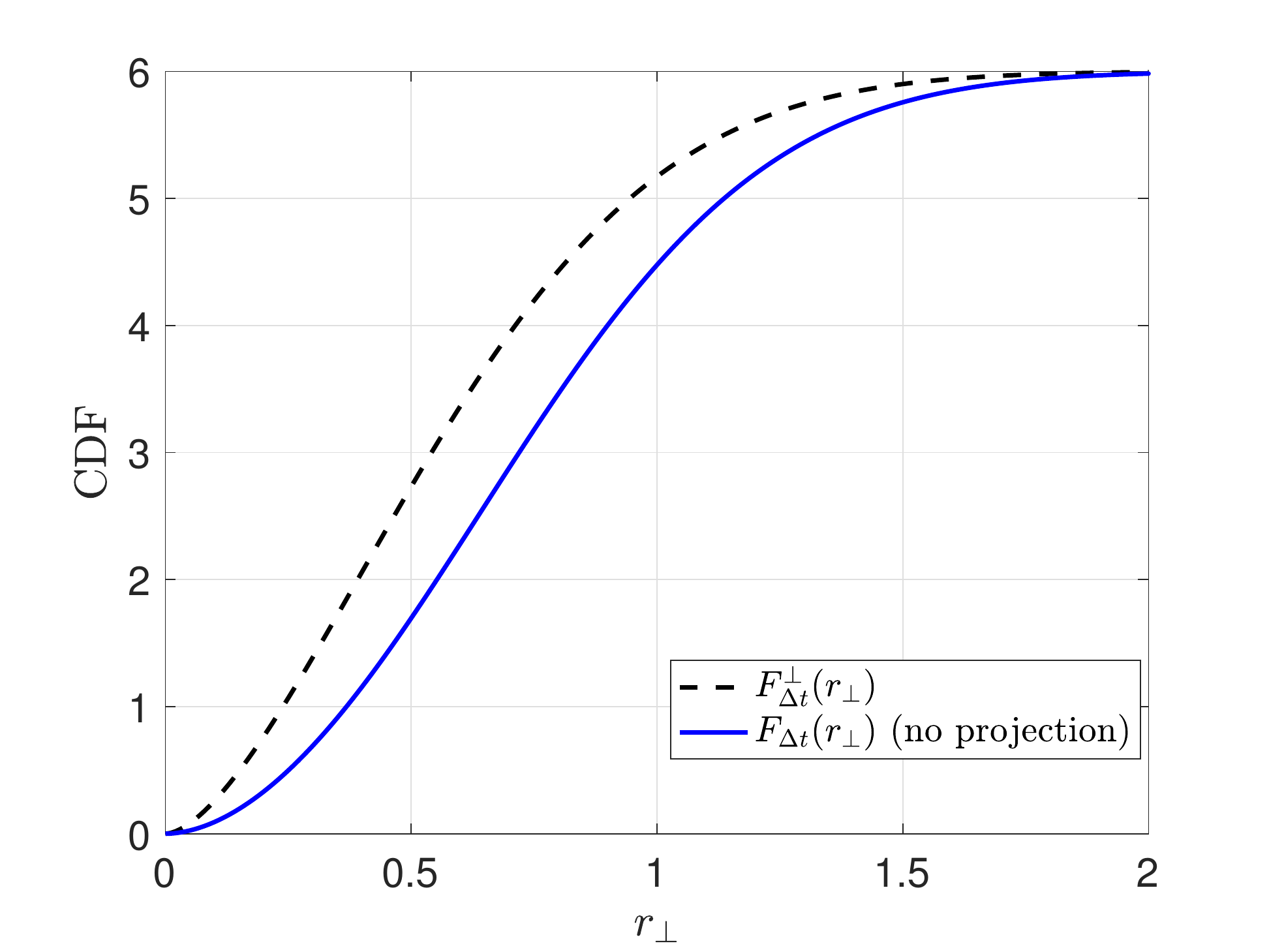}
	\caption{The effect of projection. Solid blue: unprojected radial CDF computed at the projected radius, $F_{\Delta t}(r_\perp)$. Dashed black: CDF of projected radii $F^\perp_{\Delta t}(r_\perp)$.}\label{fig:CDFproj}
\end{figure}

In reality we only know the projected distance of GCs from the center of Fornax, $r_\perp$, and not the true radial distance $r$. To obtain the CDF of projected radii, we can start with the surface density of GCs,
\be
\Sigma_{\Delta t}(r_\perp) &=& \int\limits_{-\infty}^{\infty}dz n (r) = 2 \int\limits_{r_\perp}^{\infty}dr \frac{rn  (r)}{\sqrt{r^2-r_\perp^2}} \; ,
\ee
where the 3D number density $ n(r)$ is related to the radial PDF via $n = f_{\Delta t}(r)/4\pi r^2 $. Using this relation we have
\be\label{eq:sigR}
\Sigma_{\Delta t}(r_\perp) &=& \frac{1}{2\pi }\int\limits_{r_\perp}^{\infty}dr \frac{f_{\Delta t}(r)}{r\sqrt{r^2-r_\perp^2}} \; .
\ee
The CDF in $ r_\perp $, that we define by $F_{\Delta t}^{\perp}(r_\perp)$, is given by:
\be\label{re:FperpDt}
F_{\Delta t}^{\perp}(r_\perp) &=& 2\pi \int\limits_0^{r_\perp}dR R \Sigma_{\Delta t}(R) \\
&=&F_{\Delta t}(r_\perp)+\int\limits_{r_\perp}^{\infty}drf_{\Delta t}(r)\left(1-\sqrt{1-\frac{r_\perp^2}{r^2}}\right)\;.\no
%
%
\ee
%
The CDF of projected radii contains the CDF of true radii, evaluated at $r=r_\perp$, plus another term that counts GCs at $r>r_{\perp}$ which projection casts into LOS inside of $r_\perp$. The added projection term can exceed the unprojected term, meaning that most GCs seen inside $r<r_\perp$ could be physically located at $r>r_\perp$. 
The effect is illustrated in Fig.~\ref{fig:CDFproj}.
%


\section{Exploration of initial conditions}\label{app:eccentricity}


The goal of this section is to explore the implications of uncertainties due to the line-of-sight projection in the true positions and velocities of GCs. Different projection angles lead to different true positions and velocities of a GC, affecting the orbital settling time under DF.

\begin{figure}[htbp!]
	\centering
	\includegraphics[width=0.48\textwidth]{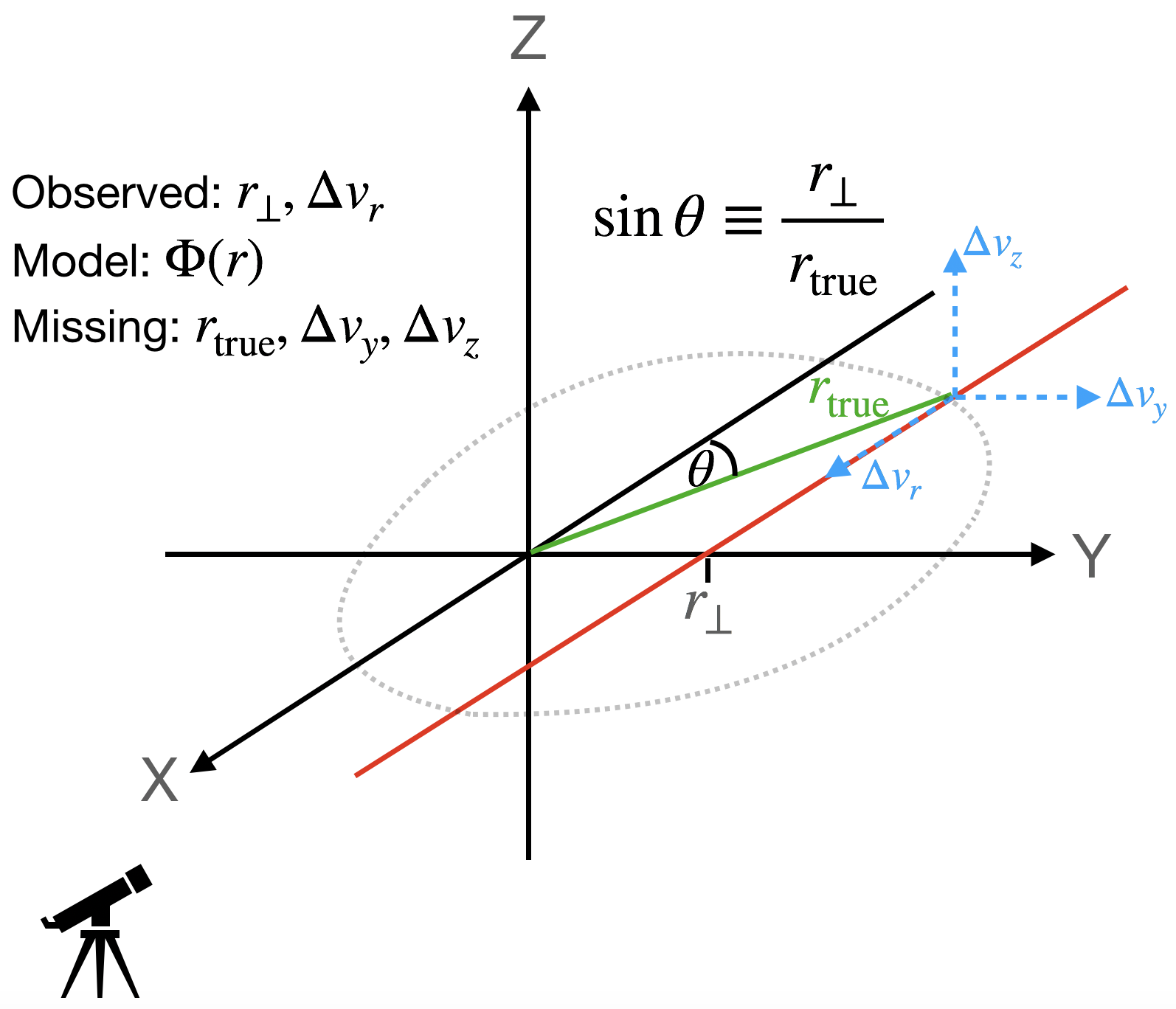}		\caption{The coordinate system that we adopt to analyze a given GC. The galactic dynamical center is in the origin. The observer is located at a very large $ X $. The GC is located somewhere on the line $ Z=0,Y=r_{\perp} $. The true radius is therefore $ r_{\rm true}=r_\perp/\sin\alpha $. We assume $ \Delta v_r $, the component of velocity in the $ X $ direction, can be measured. We assume that the rest of the components cannot be measured for now. The dotted line is the quasi-stable orbit of the GC.}\label{fig:GCsetup}
\end{figure}

Consider the orbits of test bodies in a spherically-symmetric gravitational potential $ \Phi(r) $. A given orbit lies on an orbital plane. In the coordinate system in \reffig{GCsetup}, one can parametrize the orbital plane with the unit vector $ \hat{n} = (\cos\alpha\sin\beta,\sin\alpha\sin\beta,\cos\beta) $. On the orbital plane, the radius $ r(\varphi) $ and phase $ \varphi(t) $ completely define the orbit. As in App.~\ref{app:orbits}, defining $ r=r_{\rm true} x $, $ t=T_0\bar{t}  $, $ T_0^2 = r_{\rm true}^3/GM(r_{\rm true}) $, one finds
\be
x^{\prime\prime}-x\varphi^{\prime 2} &\approx & -\frac{1}{x^2}\frac{M(r_{\rm true} x)}{M(r_{\rm true})}  \\ \left(x^2\varphi^{\prime}\right)^{\prime} & \approx &0 \label{eq:angulareomNODF}
\ee
where $ \prime $ is differentiation with respect to $ \bar{t} $. (Here we tentatively neglect DF.) The initial conditions are $ x(0)=1 $, $ \varphi(0)=0 $, $ x^{\prime}(0)=(-\Delta v_r\cos\theta+\Delta v_y\sin\theta)/(r_{\rm true}/T_0) $, $ \theta(0)=0 $ and $ |\varphi^{\prime}(0)|=\sqrt{\Delta v_z^2+\left(\Delta v_r\sin\theta+\Delta v_y\cos\theta\right)^2}/(x(0) r_{\rm true}/T_0) $. Evidently, $ x^{\prime}(0)^2+(x(0)\varphi^{\prime}(0))^2 = (\Delta v_r^2+\Delta v_y^2+\Delta v_z^2)/(r_{\rm true}/T_0)^2 $.

Given measured $ r_{\perp} $, $ \Delta v_r $ and a model $ \Phi(r) $, we explore the remaining orbital parameters which affect the inspiral time. We start with the true radius $ r_{\rm true} $. Given $ r_{\rm true} $ and, for simplicity, assuming a circular orbit, the probability of observing $r_{\perp}<x\,r_{\rm true}$ is $P(r_{\perp}/r_{\rm true}<x)= (2/\pi)\arcsin x $, because $ \sin\theta = r_\perp/r_{\rm true} $ and $ \theta $ is distributed uniformly for a circular orbit. 
Numerically, $ P(r_{\perp}/r_{\rm true}<1/2)=1/3 $. We therefore explore orbits with true radius in the range $ r_{\rm true} = r_{\perp}\times [1,2] $. 

Next, consider the velocity. We explore a total velocity in the range $ v_{\rm true} \in [{\rm max}(\Delta v_r,0.5V_{\rm circ}(r)),{\rm max}(\Delta v_r,1.5V_{\rm circ}(r))] $. Defining the eccentricity $ e\equiv (r_{\rm apo}-r_{\rm peri})/ (r_{\rm apo}+r_{\rm peri}) $, we find a maximal $ e\sim 0.2-0.5 $ for this range of $ v_{\rm true} $ in the central $ \lesssim 1$~kpc of Fornax. We note that if we decrease the lower bound of $ v_{\rm true} $ we expect smaller inspiral times. Increasing the upper bound, however, results in larger inspiral times -- but also in more tuning. A test object spends relatively little time near the pericenter. Specifically, $ T_{\rm peri}/T_{\rm apo}\equiv (r_{\rm peri}/v_{\rm peri})/(r_{\rm apo}/v_{\rm apo})\approx (1-e)^2/(1+e)^2 $, yielding about $ 1/9 $ for $ e=0.5 $. 

Considering the velocity components, we can take $ \Delta v_z>0 $ without loss of generality. The sign of $ \Delta v_y $ is, however, important: under the transformation $ \Delta v_y\to -\Delta v_y $, $ \cos\theta\to -\cos\theta $, so $ |\varphi^{\prime}(0)| $ remains constant, but $ x^{\prime}(0)\to -x^{\prime}(0) $. Since the specific energy is
\be
\epsilon \approx \frac{v^2}{2} +\Phi(r) = \frac{\dot{r}^2}{2}+\frac{l^2}{2r^2}+\Phi(r) \; ,
\ee
where $ l $ is the specific angular momentum, this transformation returns the same orbit. The inspiral time is therefore invariant under this transformation. We shall explore then $ \Delta v_y>0 $ and $ \cos\theta $ positive or negative. 

To sum up, for each GC (and a given model of the halo), we scan the range $ r_{\rm true}\in [1,2] r_{\rm proj} $. For each $ r_{\rm true} $ we scan over $ V_{\rm true} \in [0.5,1.5]V_{\rm circ}(r_{\rm true})$. For each true velocity we scan positive and negative $ \cos\theta $. Finally, we test the two cases, $ \Delta v_y= \sqrt{v_{\rm true}^2-\Delta v_r^2}, \Delta v_z=0 $ and $ \Delta v_y=0, \Delta v_z=\sqrt{v_{\rm true}^2-\Delta v_r^2} $. For each point in phase-space, we integrate the full equations of motion as in App.~\ref{app:orbits}. For each integration, we stop when $ (r_{\rm apo}+r_{\rm peri})/2 \lesssim 0.3 r_{\rm initial} $ or after $ 10 $~Gyr (the first of the two). We then denote the integration time as $ \tau_{\rm inspiral} $.

\end{appendix}

\vspace{6 pt}

\clearpage
\bibliography{ref}
\bibliographystyle{h-physrev5}

\end{document}